# “The Information as Absolute” – 2022 ed.

**Sergey V. Shevchenko[1*] and Vladimir V. Tokarevsky[2]**
**[1]Institute of Physics of NAS of Ukraine, Pr. Nauki, 46, Kiev-28, Ukraine, ret.**
https://orcid.org/0000-0003-3370-5536
**[2]Professor ret., Pr. Nauki, 46, Kiev-28, Ukraine**

**Abstract** This article presents and grounds (i.e. presents proof of the existence, the truth, the self-consistence and the completeness of) the informational concept (“the Information as Absolute” concept) in philosophy and sciences, which was developing from 2007 year till now. The concept defines the Information as ultimately common, real and fundamental “absolute” phenomenon, which exists as absolutely infinite set (“Information” Set) of elements (members) and informational (e.g., logical) links between the elements; where any element itself is some informational structure also. Correspondingly, for example, Matter as the substance, radiation, etc., is some system of informational patterns, constituting a specific, and practically infinitesimal comparing to the Set, element “Matter” of the “Information” Set. The concept allows for the resolution, or at least for a consideration on a higher level of rational comprehension, of basic ontological and epistemological problems in philosophy and natural sciences; it clarifies basic fundamental notions such as space, time, energy, etc., and so is the fundamental base for real development of science.

## Content

## 1. Introduction

Yet in Ancient times, two main ontological (and correspondingly, epistemological) philosophical concepts were formed — Materialism and Idealism. Both concepts were, and still are, based on really only transcendent beliefs in some transcendent fundamental Essences. In Materialism such Essence is some for some transcendent reason existent, eternal and eternally moving, omniscient and omnipotent "Matter", when everything what humans observe, including themselves, is some realization of "Matter", while consciousness is the "highest level of Matter's development"; in Idealism a number of (also transcendent, eternal, etc.) Essences are considered – "Consciousnesses", "Spirits", "Ideas", etc., when everything what humans observe is some "emanations" of "Consciousnesses", "Spirits", etc., while Matter is the "lowest level of the emanations", etc.

As both concepts are no more than transcendent beliefs, and so it is fundamentally impossible to prove or disprove the truth of any of them, though corresponding attempts, discussions, disputes (sometimes rather radical) took place over and over again within rather long time, and so both opposite doctrines exist till now equally legitimately in the mainstream.

But in reality the problem of the transcendence of the Essences above, as well as of others really transcendent in the mainstream fundamental phenomena/notions, is resolvable because indeed ultimately fundamental Essence, which is the base of everything, namely Information, isn't transcendent and can be, in principle, studied. The substantiation of corresponding informational ("the Information as Absolute") concept in philosophy and other sciences, first of all in physics and consciousness sciences, which was developing from 2007 year till now [Shevchenko and Tokarevsky, (2007-2008), (2008-2010), (2008-2010), (2013-2017)], is presented in this article.

The concept defines Information as the ultimately common, *absolutely real and absolutely fundamental* concept/phenomenon, an "Absolute", which absolutely fundamentally really exists, and is actualized as the absolutely infinite set ("Information" Set) of informational patterns/systems — elements (members) of the Set, and informational (e.g., logical) links between the elements, where every link is some informational structure also. Correspondingly, for example, Matter as the substance, the radiation, etc., as well as Consciousness, are some actualizations of Information, which exist as specific and practically infinitesimal comparing to the Set, elements of the Set.

The concept allows for the resolution or at least for a consideration on a higher level of rational comprehension, of the basic ontological and epistemological problems in philosophy and fundamental problems in natural sciences.

Finally note here, that because of the principal transcendence of all fundamental phenomena/notions above in the both main mainstream philosophical doctrines — and numerous sub-doctrines, schools, etc., really numerous publications about ontology of "Matter" and "Consciousness", "Space", "Time", etc., are nothing else than some really transcendent constructions, considering of which would have only some historical interest, so this paper addresses to some points and publications in the mainstream philosophy only in some specific cases.

## 2. On the concept of "Information"

It is rather interesting that the discussion "so what is the information?" in scientific, technical and philosophical literature goes on in many years already without any consistent results. [Abdeev/ Абдеев, 1994]:

> "Depending on a branch of science where an investigation was carried out, information got a large number of definitions: information is an indication of a content, obtained from external world in the process of adaptation to the world (Wiener); information is a negation of the entropy (Brillouin); information is the communication resulting in a decreasing of an uncertainty (Shannon); information is a transmitting of a diversity (Ashby); information is an originality, novelty; information is the measure of a structure's complexity (Moll); information is a probability of a choice (Yaglom); etc. Each of these definitions reveals one or another aspect of this polysemantic concept".

Here is no room for a detailed analysis of this discussion, we note only that its productivity turned out to be rather poor, from what follows, for example, large number of existent definitions of information. Chernavsky [Chernavsky/ Чернавский, 2001] gives more then twenty different ones. Capurro and Hjørland [Capurro and Hjørland, 2003] quoted some dissertation where about 700 definitions were collected.

Let's consider some of the definitions (mainly cited in [Abdeev/ Абдеев, 1994]) that have essential semantic distinctions:

1. (Philosophical *encyclopedia*) "Information (lat. "informatio" – an examination, a notion, a concept): 1) a report, a notification about a state of affairs or about something else that is transmitted by a person; 2) decreased, removed uncertainty as a result of the communication obtained; 3) a notation inherently relating to a control; signals and their syntactic, semantic and pragmatic parameters; 4) transmission, reflection of the variety of any objects and processes (of alive and non-alive nature)".

2. "Information means some order, a communication, creation of the order from a disorder or, at least, increased regulation compared to that which existed before the communication was obtained".

3. "Information is the manifestation of the property of the objects of living nature to reflect in the form of some mental sensations the movement of the objects in surrounding world".

4. "Information… is a quality of the objects, phenomena, processes in the objective reality and of man-made controllers, which lies in the ability to conceive an internal state as well as the state and the impacts of an environment and to preserve, sometimes, the results; to transmit the data about the internal state and cumulative data to other objects, phenomena, processes".

5. "Information is a philosophical category that is considered along with such as Space, Time and Matter. In the most common form information can be presented as a notation, i.e. a form of some relations between a source which communicates and a receiver which obtains a notation".

6. "Information, as well as matter, exists and has always existed… information is some integral attribute of Matter and movement which realizes a certain way of Matter existence and presents some measure of the changes which follow all processes occurring in the World".

7. “The phenomenon of information is a multi-stage, irreversible process of coming into being of a structure in some open imbalanced system that begins at a random memorized choice which this system carries out when it transforms from chaos to an order, so the process is completed with a purposeful action according to an algorithm or program that are in accordance with the semantics of the choice.” [Melik-Gaikaz’an/ Мелик-Гайказян, 1998].

8. “Information is some qualitative and quantitative characteristic of the level of reflection. Generally information is a quasi-force which is directed against disorder and chaos; in this sense it cannot be separated from structure and regularity of material systems” [Berg et Spirkin/ Берг и Спиркин, 1979].

9. (Weizsäcker, 1959, quoted in [Yankov, 1979: 39]) “Now many people begin to recognize that it is necessary to consider information as something third that differs from Matter and consciousness… This is Plato’s Idea, Aristotelian Form, invested by such a way that the human of XX century assumes to know something new from it”.

10. [Wiener, 1983] "Information is information, not Matter or energy. No materialism which does not admit this can survive nowadays”.

11. [Landauer, 1999] “…Information is inevitably inscribed in a physical medium. It is not an abstract entity. It can be denoted by a hole in a punched card, by the orientation of a nuclear spin, or by the pulses transmitted by a neuron”,

- and, at last

12. “…If you are interested in the question “what is information?” and find corresponding definition in some book (which is, generally speaking, rather difficult, since the authors usually keep from giving such a definition), then in great likelihood other authors will not agree with this definition.” [Petrushenko/ Петрушенко,1971].

It seems quite natural that the last author possibly had some grounds for such evident pessimism. However, as will be shown below, in reality the problem of the definition of the concept/ notion “information” can be solved, or at least can be evaluated in the general way, by using logical analysis.

Besides, note that all listed definitions have a common conceptual flaw, each of them is tautological: “information is information” (or “data”, “algorithm”, “communication”, “evidence”, etc.) Thus any attempts to define the concept/ notion “information” through something, which is more common and fundamental, turn out to be ineffective, whereas now in textbooks one can find a number of “information theories” such as Shannon’s theory, a number of complexity theories, theories of algorithms and automata, etc.

## 3. On the notion “the set”

Another fundamental notion that will be necessary to build this informational concept is the notion “set”. It turns out that in attempts to define this notion in mathematics the same problem as at defining of information arises, since any definition becomes a tautology: the set is the set, ensemble, manifold, collection [of the elements], etc. The difference is practically only in that the mathematics has been evolving by way of maximal formalization and using rigorous logical rules/ limitations at creation of a next domains of this science, whereas the attempts to formalize concepts/ elements/ concatenations in the information theory were essentially lesser productive.

Now in a number of the set theories the notion of a "set" is taken as an *undefined primitive*, which can be defined only *restrictedly*, i.e. by *defining its properties in a limited system of axioms*. Though there are some set theories where the notions of the set are defined "completely" (e.g. [Vavilov/ Вавилов, 2007]) as well as the theories where some "more common" [relating to the set] notions are used, for example the notions of the categories and the toposes [Goldblatt, 1979]; [Baez, 1999]; [Marquis, 2003]. But such notions are only certain (sometimes not natural) natural extensions of classical G. Cantor's definition: "Unter einer Menge verstehen wir jede Zusammenfassung *M* von bestimmten wohlunterschiedenen Objekten in unserer Anschauung oder unseres Denkens (welche die Elemente von *M* genannt werden) zu einem ganzen" ("By a "set" we mean any collection *M* in a whole of definite, distinct objects *m* (which are called the "elements" of *M*) of our perception or of our thought").

## 4. The relations of information and set

So in mathematics, there is a number of the information and set theories when the corresponding notions aren't, in fact, defined.

To clear the problem let us recall the Cantor's definition of a set. In this definition the key is "*of definite, distinct* objects … of our perception or of our thought" i.e. to define a set turns out to be impossible without notions (terms) which relate to the notion "information". And, in turn, information appears if and only if alternative of some elements (that is, diversity) of some set appears. I.e. the system "a set + an information" exists always as a unity, the *set is a form (a mode) of existence of the information.* The notion "set" here, naturally, is used in a broad sense, i.e. not only as a "collection of some elements". On a set any informational connections (e.g., mathematical operations) between the elements can/ should be defined (see the definitions of the information above, definitions of the categories, the toposes, etc.) which define the set's (and the set's elements') specific properties by establishing a system of axioms.

It is already well known that complete set-theoretic axiomatic system is, very probably, infinite, and now we can conclude that the same inference is true for the informational theory. Nevertheless, recognizing the unity between the concepts of set and information allows us to build rather general and effective approach at further consideration of this informational concept.

## 5. Some properties of Information

As was already mentioned, the notion "information" unlike the notion "set" is essentially less formalized; a rather poor system of axioms exists for the information. Current formalized theories – Shannon's (applications in the communication theory and physics), theories of complexity, algorithms, and automata (cybernetics) – reflect (allow to formalize) the properties of this phenomenon/notion only restrictedly. Such a situation follows from both infinite complexity of this notion and limited capability of the languages, including limited capability of individual (human's) interpretation of the words/ notions. Nevertheless, we can formulate a number of common basic properties of the information in addition to the "definitions of information" in Sec.2 above, which, in fact, define only some certain specific properties of information.

**Property I1**. *The phenomenon "Information" is objective and doesn't require existence of any "sentient being" to exist.*

**Property I2**. *Information can exist at least in two possible modes: 1) "fixed information"*, e.g. a picture, a computer code listing, and 2) *"dynamic information"*, a changing picture, an execution of a program code in computer, moving particle, etc.

Here we should make some "epistemological" remark. For further consideration, note that any *indeed new* information about the external [to a human] World can be obtained by a human's consciousness only as a result of some experiments, *any indeed new knowledge is empirical.* This new knowledge in a science becomes "axiom(s)", "postulate(s)", "Nature law(s)". Further, a human consciousness applies the axioms for more detailed analysis of specific natural processes, e.g. creating theories or solving technical tasks, mathematical problems; etc.

However, rigorously speaking any empiric postulates, etc., "have no right to be laws". In reality they always remain hypothetical, since they are based on the necessary, but principally insufficient, criterion of the reiteration of given experimental outcomes in given experimental conditions. From the fact that in *N* experiments some identical (in fact, nearly identical) outcomes were obtained, by no means follows that the outcome in (*N*+1)-th experiment will yield the same. Logically, say a physicist, can only believe in that the next result will be "in accordance with the theory". For example, well-known Newton's statement "I do not feign hypotheses" is incorrect, and, e.g., Newton's gravity law (as well as any other Nature law, though) is no more than a hypothesis, though claimed as the "law" postulate in physics.

Moreover, as it was proven by K. Gödel [Gödel, 1931], it turns out to be that there exists some limit for the complexity of a mathematical theory when the theory based on a consistent system of axioms becomes incomplete, i.e. when there are some true statements / propositions which cannot be proven in that theory. An example, possibly, is the fact of non-provability of the "continuum hypothesis" in Zermelo-Fraenkel set theory that was proven by Gödel and Cohen [Gödel, 1940]; [Cohen, 1963].

The pointed above (the definitions 1-9, 11,12 in section 2, properties **I1, I2**) properties of the information, if claimed as some "postulates", *are some empirical assertions* also and in this sense these postulates by any means don't differ from, e.g., Newton's gravity law. However, there is the *fundamental difference* between the information's postulates and the postulates in Nature sciences.

In the case of Information, we have *fundamentally another situation. It is sufficient only once* to "discover in an experiment" an information *as a data*, then at once it can be logically proven that for the information a number of basic properties and rules are *always rigorously and completely true*.

As the *main* **Property I3** is true, which we obtain as follows. Let us consider the notion of a "null (empty) set" that is introduced in any set theory: a null set is the set that contains no members/elements (e.g. [Hrbacek, Jech, 1999]). This set, unlike any other sets, is unique – null set exists as the single set, irrelatively of how many and whatever sets exist anywhere (at that sometimes it is possible and useful to introduce the specific empty sets for some specific sets, though). And further, if we recall that any set is, generally speaking, a mode of existence of some information, then we must conclude that the null set contains all/ any elements of all/ any sets. Indeed, to define the null set is necessary to point out that this set doesn't contain this, this,

this… and so on, down to "absolute" (the term "absolute" will be correctly defined below in this section, Property **I6**) infinity, element of all/every sets – or of the set of all sets and elements); it turns out to be that the null set isn't so empty as it is adopted in mathematics.

The notion "null set" in the "informational" language one can formulate as the statement *"there is no anything*" (or "there is nothing"). And just as that was in the case of the null set's notion, we can conclude that the statement *"there is no anything" contains complete information about everything about what exists, what can exist (as well as about what "cannot exist", but exists as a false information) in the absolutely infinite set of informational patterns/systems of the patterns, which we call here the Set "Information",* which contains complete information ("data") about everything.

However, it is necessary to make an evident revision of this statement, since it is incorrect, as there exists the information that there is no anything. Correspondingly true will be *infinite cyclic statement ("Zero statement"): "there is no anything besides the information that there is no anything besides the information…"*. I.e. Zero statement is at the same time fixed and dynamic informational pattern.

From the above two important consequences follow (i) – *any information absolutely for sure cannot be non-existen*t, and so (ii) – the "Information" Set exists absolutely always, "in absolutely long time interval", absolutely fundamentally having no Beginning and no End.

Note, though, that in the above the fundamentally always existent information about everything exists in two states – "real explicit" informational patterns/systems, and "implicit" information – which is indicated in the negations. An example: the Zero statement is real informational pattern, whereas its absolutely infinite content about all other the Set's elements "no this, this…." element of the Set is in the implicit negation statement.

Let's return to the definitions 1-12 (except, of course, Wiener's one) in section 2 above. Most of these definitions contain tacit assumption that for an existence of an information some storage device is necessary, a brain (e.g. a human's one), papyrus, computer, some thing with observable properties, etc. However, Zero statement containing absolutely infinite information exists when, by its definition, there are, including, no storage devices. From this follows:

**Property I4**. *For the existence of information there is no necessity in the existence of an external storage device, but since some storage device is, nevertheless, necessary, then only one possibility remains – when information itself is a storage device of information.* Though this implication could have been obtained earlier from the "experimental fact" that any definition of information appears to be a tautology: the facts that information can be defined only via information itself and that information is itself a storage device are, practically, the same. Note, though, that in this case all/every elements in the Set have, nonetheless, specific "storage devices" – concrete "spaces", which are actualizations of the absolutely fundamental phenomenon/notion "Space", which (concrete spaces) are parts of the whole Set's space (more see section 6.1. below)

Carrying out analysis similarly as it was in the case of Zero statement again, we obtain

**Property I5**. *Any element of the Set "Information" contains all/every elements of the Set, i.e. any element contains the Set "Information" totally.* Indeed, to characterize (single out) some element from the Set, it is necessary to point out

all/ any distinctions of this element from any other element; every element in the Set exists as a bit "I/not-I", where the part "I" is real information about the element in its current state, and "not-I" section contains complete implicit information about all/every other elements of the Set, including – about given element in other times of its existence. Correspondingly, note here that the Set, in contrast to sets in mathematics, so has no null set, the Set's element informational pattern "Zero statement" doesn't differ principally from any other element.

The list of information's properties is infinite, but even the properties **I1**- **I5** convincingly show the originality and fundamental nature of the information's concept/notion. Besides, from these properties follows:

- (independent on anything) ***existence*** of absolutely infinite and fundamental "Information" Set, as well as of introduced here informational concept;

- ***completeness*** of the informational concept, since in the "Information" Set exists no conceivable operation when some element could quit the Set. Besides, the Set contains all/ any possible false information. And its amount possibly infinitely exceeds the amount of true information, though when we deal with "absolute" infinities, such a statement possibly requires some separate study;

- (self-) ***consistence*** of this informational concept. Indeed, the consistence of some theory/concept in mathematics implies that in this theory it is impossible to prove truth of (at least two) logically inconsistent implications, because one of the implications must necessarily be false. In other case the theory is inconsistent and therefore false. In the case of this concept such an interpretation becomes inapplicable, including because obtaining false information does not lead out it from the Set;

- because of absolute *completeness* of the information concept we principally cannot go out of the concept in order to prove it's (and the Set's) ***uniqueness***.

Note, also, some other basic properties of the information:

**Property I6**. *Since a process of transformation (e.g. determination) of some specific information reduces to enumeration of variants, the "Information" Set, in spite of its absolute infinity, is, very probably, at least in very certain sense, discrete.*

Property **I6** (and the text above) contains at least two notions that call for additional explanation. First is the notion "*discrete*" applied here (though with a stipulation "very probably in very certain sense") to the Set totally, when there is, e.g., the notion of the continuum (continuum is, of course, a subset of the Set), which is by definition non-discrete. Secondly, in standard set theories it is often accepted that the "absolute infinite" set, or "set of all sets" doesn't exists. If one assumes that such a set, $X$, exists, then it is possible to create power set of this set, $2^X$, and the cardinality of the second set rigorously exceeds the cardinality of the set $X$. However it is known, that if the continuum hypothesis is true, then the cardinalities of the continuum set and of the "discrete" power set of the natural numbers set, $\aleph$, i.e. $2^{\aleph}$, are equal, so the continuous and the discrete are in certain sense equivalent. Thus, e.g., infinity sequence of power sets for, e.g., natural number set: $Y_0=2^{\aleph}$,… $Y_k=2^{Y_{k-1}}$ …, $k\rightarrow\infty_A$ (when $\infty_A$ means, in turn, "absolute infinite"), must have maximal cardinality (be "absolute infinite"), including since in this case the concept of a concrete "next power set" loses sense above.

Another approach at the consideration of the "absolute infinite" problem follows from the [arithmetic] zero's notion. Though practically any arithmetic contains “0” and in most cases zero is used as some number, it is not an usual number. It is introduced in arithmetics always by some additional rules. In reality zero indeed *is not a number*, it is the specific “numbers” empty set: “there is no numbers”. Therefore, in arithmetic it is permitted the division of any quantity by any infinitesimal quantity, but the division by zero is prohibited, since its result is “absolutely uncertain”, and so, say, by using formal division on zero in mathematics it is possible “to prove” the equality of any different arbitrary numbers, say, 1=2.

Here we can say that the “absolutely infinity” can be (as an axiom at the definition?) considered as the result of the division of some “usual number/quantity” by the zero. Though if such definition would be introduced in mathematics, in this case many things in mathematics would turn out to be rather mathematically strange, as the above, however really, again, every element in the Set contains whole Set, and so really all the Set’s elements are “equal”, including 1=2=the Set.

**Property I7.** *(At least true) information in the “Information” Set, as well as in any of the Set’s limited (by some attributes) subsets, can be “absolutely exact”.* For example, two identical texts contain absolutely exactly identical implications.

**Property I8.** *From that Zero statement, which contains all data about everything, is expressible in practically any human’s language rather possibly follows that any information from the “Information” Set can be expressible in practically any language.*

If this language (or maybe more correctly, if a corresponding consciousness is capable) is capable for infinite development, though.

Finally note here, that though we cannot to define the notion/phenomenon “Information” because of its ultimate fundamentality through some more general and fundamental notions, and really cannot define it by defining properties - the number of the properties is practically infinite, however *the phenomenon “Information” isn’t transcendent*, and so has some “operational” definition”

“***Information is something that is constructed in accordance with the set/system of absolutely fundamental Rules, Possibilities, Quantities, etc. — the set/system “Logos” in this concept***”.

Or, by other words, the “Logos” set elements “make something to be information”. Note, though, that something like this approach was seems firstly intuitively applied by Aristotle, when he really attempted to analyze the phenomenon “Information” and developed his *categories* system, and some of his categories are some “Logos” set elements. Note also, that the attempts to define some ultimately common elements in Information were made a number of times later, so now there exist a number of categories and “*universals*” systems, some of which are some “Logos” elements as well; and in this concept the “Logos” set elements aren’t defined completely, that is one of subjects for study in philosophy at development of this concept. However, a few of the “Logos” elements are known, and are utmost important, and so these elements are more in detail considered here.

## 6. Some selected elements of the “Logos” set

The “Logos” set elements are “absolutely fundamental”- unlike the phenomena/notions “Matter” and “Consciousness”, which are fundamental in the mainstream philosophy and sciences, while are only some concrete practically infinitesimal elements of the Set, and so are "simply fundamental"[1]

### 6.1 The elements "Space" and "Time"

Space and Time are defined in encyclopedia as some "universal forms of Being of Matter, its prime attributes", which characterize "extension/ length" and "duration" of the Being. It is rather easy to note that these definitions contain some evident flaws. The concept "Space" is defined through, rigorously speaking, non-defined concept "extension", "Time" is similarly explained by "duration". Though the concepts of the extension and of the duration can be, to some extent concretely, determined empirically, the same questions remain: from where/how did these "forms and attributes of Being" appear?

In the concept "*Space*" and "*Time*" are absolutely fundamental *Rules/Possibilities* that are absolutely fundamentally necessary for any informational pattern/system could exist:

- "*Space*" is necessary for any information *could exist at all*, and

- "*Time*", *additionally to Space*, is necessary for some informational pattern/system could be dynamic, i.e. could change.

"Space" as the *Possibility* makes be possible placing of concrete informational patterns/systems in concrete "*space*", which (the space) at that is realized as a concrete set of "*space dimensions*", which (dimensions) are *necessary to actualize independent degrees of freedom* of the concrete patterns/systems at changing of all their possible states.

Since Space is a logical possibility, the sets of the dimensions form so concrete, and principally infinite, "empty space containers" for the concrete one type patterns/systems. For a space it is all the same – how many one type patterns/systems, which are constructed by the same concrete sets of logical rules/links/constants, and so have the same degrees of freedom at construction and changes, are placed in the container.

And it is all the same – in what places in the infinite container the patterns/systems are placed. The unique requirement, when Space acts as the *Rule* is that a non-zero "space interval" must divide the different patterns/systems, and any pattern/system must occupy non-zero "space interval" (non-zero "space volume", if there are more than one intervals in different dimensions) as well.

Since any information absolutely fundamentally cannot be non-existent, everything had happened/existed in the "Information" Set; and everything is happening/existing, and will happen/exist always;

- and the concrete patterns/systems, including Matter and consciousness, simply use the fundamentally always existent concrete spatial dimensions from the at least "simply" infinite "number" of always existing space dimensions of the Set's whole space in concrete actualization of current state of concrete pattern/system. As that is, for example, for Matter and humans in this concrete actualization of this Universe evolution.

[1] This division on "absolutely fundamental", when a notion relates directly to the phenomenon/notion "Information" and "Information" Set, and [simply] "fundamental", if relates to fundamental only for humans phenomena/notions, problems, etc., is used in whole text

To define Time there are a lot of approaches now, up to the statement that Time does not exist (see, e.g., [Rovelli, 2009]). J. A. Wheeler [Wheeler, 1986] wrote about Time in a similar way as in encyclopedia:

> "...But time: how is time to be reduced to more primitive concepts? Explain time? Not without explaining existence.... Explain existence? Not without explaining time. To uncover the deep and hidden connection between time and existence ... is a task for the future..."

Really "Time" as the *Possibility* in main traits is analogue to Space, Time is "the space for changing states of changing patterns/systems", and exists/acts in concrete cases forming, including, corresponding "time dimension" for dynamical patterns/systems.

However, Time has the essential difference from Space: for Time it is all the same for/by what reason/way, by what degree of what freedom, etc., and in what informational pattern/system a change happened.

So in this case it is enough to have only one absolutely fundamental and universal time dimension, which exists and acts in whole "Information" Set for all changing states of all dynamic the Set's elements, and so is the *time dimension* in all/every concrete *spacetimes* of concrete patterns/systems.

Time as the *Rule* also acts as that a non-zero "time interval" must be between different states of changing patterns/systems. However, in this case this Rule, unlike Space one, seem as is determined by a couple of two, on first glance different, absolutely fundamental and "external to time" causes. The first one is that any information if appeared can not be non-existent, and so the next changing state can not "erase" previous state, so must be placed in some other point in corresponding concrete spacetime. The second is that a continuous changing of states is impossible, because of the logical self-inconsistence of the Logos set element "Change" (more see below), and the changes happen only along non-zero (discrete) time intervals.

At any change of any informational pattern/system this pattern/system moves in the time dimension on corresponding time interval $\Delta t$, in every case, when the changing pattern/system is fixed in space, and at every change of its spatial position on, say, $\Delta x$. At that the changing of a pattern/system spatial position can be in principally arbitrary number of space dimensions, whereas all dynamic elements in the Set move at every change only in one, universal time dimension.

Finally, in this section we make a brief remark to existent definition of "Time" in recent philosophy and physics. This definition was firstly done by [Newton 1686]

> "*...Absolute, true and mathematical time, of itself, and from its own nature flows equably without regard to anything external, and by another name is called duration ....*"

- at that for Newton, correspondingly, clocks show the time flow independently on time and only because of they also tick equally equably.

This definition, however with the two relativistic modifications, remains in recent sciences. According to special relativity postulates time (i) - not only flows equably, this flow depends on motion, and, whereas in stationary inertial reference frame time flows in accordance with Newton's definition, in moving frames its flow becomes be dilated, and (ii) – time governs material bodies, including clocks, and so "time is what clocks read", and clocks show in stationary frames "Newton's" flow, and in moving frames – dilated flow. Besides this time flow is observed as an "arrow of time" [Eddington, 1948].

From the correct definition of "Time" above it follows that there cannot be any, "Newton's", "normal", "dilated", etc., "time flows", and any "arrows of time" as well – and fundamentally time cannot impact on anything. Matter, and every material object/system, including clocks, simply constantly, because of the energy conservation law, are changing, and so move in the fixed, of course, time dimension, passing from given states to mostly more probable states; when a changing is deterministic, that only connotes, that the probability is equal to 1. Clocks are special material objects that – rather specifically, though (more see [Shevchenko, Tokarevsky, 2021], show how they move in the time dimension.

### 6.2. The element "Energy"

Note here that as the *Rules* "Space" and "Time" are real grammar rules in every language, and most of other "Logos" elements rather clearly relate to intuitive commonplace — unlike "Space" and "Time" — perception of the phenomenon "Information", some examples that are the phenomena/notions – "Logos" set elements - "Change", "Causality", "Logical Rules", "Relation", etc.

Unlike the "Logos" set elements above, Quantity "*Energy*" is rather specific element, which doesn't relate to the basic properties of Information in "Information as a data", and so at least on first glance it looks as something that is beyond Information. However, it is absolutely fundamentally necessary for to change, including, of course, to create, of any/every real informational pattern/system. That is because of the fundamental logical self-inconsistence of the other absolutely fundamental phenomenon/notion, "Logos" set element, "Change":

- at every change of something its state is simultaneously former, recent, and future states, when all the states are different by definition. That is logical nonsense.

To overcome this logical prohibition of changes at every change it is necessary to pay by two points:

(i) – to change [including to create] some informational pattern/system it is necessary to spend some non-zero portion of "Energy". However, that is not enough if the portion is finite; and so, besides,

(ii) – really at any change the changing state on some level/scale is uncertain – "illogical".

Including just by this way in Matter quantum objects/events/effects/processes exist and happen.

Note, though, that the fact of impossibility of deterministic continuous changes of anything was proven more 2500 years ago by Zeno in his brilliant aporias (some really questionable analysis of the aporias see [Dowden, 2010]), when Zeno, in fact, predicted the quantum mechanics. Achilles overcomes a turtle because of $\Delta V \Delta x \geq \frac{\hbar}{2m}$, and when Achilles's $\Delta x$ and a turtle's $\Delta x$ are such, that Achilles illogically occurs ahead the turtle, Achilles further runs without any logical problems. Relating to QM, note also here, that in this presented concept directly it follows the answer to the fundamental physical problem:

“*Why does the QM postulate exist and is adequate to the reality that all given type particles are identical, and why is it adequate to the reality*” — this QM postulate is adequate to the reality because all given type particles are copies of the corresponding unique informational patterns, that is a typical situation in Information, see **Property I7** above.

So, remembering that everything in the “Information” Set, including our Universe before Matter’s creation, always, in absolutely infinite time interval, exists in two main — “real” and “implicit” states, “Energy” is necessary, and is always used, to open the cans “there is no something”, and to make this something to be real.

However, again, Energy looks as something that doesn’t directly relate to the absolutely fundamental and extremely complex, nonetheless principally cognizable, phenomenon “Information”, including practically all other “Logos” set elements, and so Energy till now — unlike all really fundamental phenomena/notions, which are principally transcendent in the mainstream philosophy and sciences, but have become to be non-transcendent in this concept — remains to be a mysterious element of “Logos” set in this concept as well.

And, though in the concept it is essentially clarified how Energy acts in concrete simple logical system “Matter” – see above and the informational physical model [Shevchenko and Tokarevsky, 2021], which is based on the concept, however Energy remains, including in Matter case, now completely mysterious - so from where and how some energy appeared to create Matter? — and to create the other, also fundamental, elements in the Set “Consciousnesses”- we don’t have in mind the transcendent “Consciousness”, “Spirts”, “Ideas”, etc. in philosophical Idealism doctrine, but the real informational systems, versions of one of which are observable on Earth as consciousnesses of living beings

Though, besides the above (for which Energy is necessary), now it is also understandable that Energy is rather “dull” Quantity, and the changes in informational patterns/systems are eventually determined by concrete information in/of concrete changing/creating patterns/systems.

However, that till now is not too essential in physics. The reason is that Matter is rather simple logical system, which is based on a limited set of fundamental and universal basic logical rules/laws, links, and constants (more see below), where the exchange by energy at material objects interactions is, in depth, highly standardized and universal, and the dependence of the action of Energy on difference of informational content in different material objects so is inessential, besides that there are, correspondingly, a few “forms of energy” – “kinetic”, “thermal”, “nuclear”, etc.;

- and, if we don’t address to the question “from where and how energy in Matter appeared at Matter’s appearance”, this problem isn’t actual because of the energy conservation law action in Matter, and at Matter’s constant evolution after Creation only some redistribution of the primary energy portion proceeds.

Besides note here, that actualizations of Energy action are as a rule concretized as that relate to concrete changes of states of some informational patterns/systems in accordance with what concrete degree of freedom of the changes is actualized. In this case other absolutely fundamental “Logos” Quantity also acts – “Momentum”, which is directed in informational patterns/systems’ “spaces”, and in this case the fundamental uncertainty of Change above reveals itself as “momentum uncertainty", including whereas energy of the patterns/systems in some cases doesn't change.

And what looks just as real, there exists a more fundamental and mighty than Energy phenomenon: "Logics" — and Information itself also, correspondingly. Though Energy on first glance seem as something external to Information, for example to some data, nonetheless it evidently could exist the *real* state, when there is "nothing", including "no energy", nonetheless in this case there principally exists — since logically cannot be non-existent, the real "Zero statement" in the concept.

### 6.3. The element "Inertia"

Inertia, correspondingly, is absolutely fundamental phenomenon that characterizes the logical resistance to changes because of the self-inconsistence of "Change" above. As energy, the inertia in simple informational system "Matter" can be, and is, characterized; according to Newton, by the physical parameter "inertial mass". Note here, that that has no relation to the existent in standard physics explanation of what is the inertial mass as some action of the Higgs field.

On an aside, note a tenet, rather popular in official physics, that "energy and mass are two faces of one coin, one of them converts to another". That is fundamentally incorrect. Both absolutely fundamental phenomena "Energy" and "Inertia" indeed absolutely fundamentally always co-exist in every informational pattern/system, including in every material object, but they are fundamentally different, and so, say, at the interactions in Matter first of all energy transforms/is distributed into energy, though with obligatory accompanying by transformation/distribution of inertial masses.

Finally note here again, that both — Energy and Inertia always absolutely fundamentally obligatorily co-exist in all/every dynamical elements in the Set, including at existence and constant operating of the informational system "Consciousness" – to think it is necessary to eat; and some mental constructions in head sometimes are observed as well inertial.

## 7. Application of the concept: Matter, Consciousness, AI, Life

It seems rather evident that "What is Matter?", "What is Consciousness?", "From where (how) did they appear to be?" are main questions in main mainstream philosophical branches — ontology and epistemology. Under *necessarily empirical* (see section 5) approach, which a human's consciousness applies to perceive the External, because of real fundamentality of both phenomena/notions it is impossible in this case to the obtain rational answers on these questions. The evidence for such a conclusion is longtime co-existence of two main competitive opposite philosophical concepts — and doctrines, Materialism and Idealism. Both doctrines have held the in fact futile dispute for a number of thousands years, and this long experiment practically unambiguously shows that both doctrines are nothing but transcendent beliefs, it is impossible to prove, or even to test empirically, the truth of any of them.

Materialism's foundation is "the system of Nature laws"; however, as that was pointed out above, any Nature law is essentially empirical and so can only be postulated, in other words, be taken without a proof as something fundamental, while in this case yet next question appears – some laws seem as very probably exist – however for what non-mystic reason does that happen? This question by no means has some really non-transcendent sense in Materialism, which really is nothing else than a transcendent belief in the Great Materialistic Principle "That is so because of that is so". As well as Materialism fundamentally is not capable of answering to the main epistemological questions – why and how humans study Nature at all? and why sometimes results of some studying are adequate to the reality? Note, though, that answers to any really fundamental epistemological question is possible only if the ontological problems "what is "Matter" and "what is "Consciousness" are scientifically clarified, and so existence of the branch "epistemology" in the mainstream philosophy looks as that has no essential grounds.

Idealism is more rationally grounded; it states that a sentient Creator established Nature laws when He created this Nature. However, as early as in 18th century I. Kant [Kant, 1787] showed that it is impossible to prove the existence/ non- existence of the Creator. Besides, to create Nature "from nothing" Creator must be omnipotent, when, as it was proven yet in Middle Ages, any omnipotent being is logically self-contradictory. Correspondingly in Idealism some "materialistic" questions appear, for example from where and how the Creator happened to be? And, of course, why and how Nature is as it is? — these, and not only, questions have in Idealism also only some transcendent, and, correspondingly numerous in rather numerous sub-doctrines, schools, etc., "answers" – as that is in Materialism, of course, also.

Presented here informational concept allows to clear up the situation essentially. As it was proven above, any information exists fundamentally absolutely always, or "in an absolutely infinite long time"; it fundamentally, logically, cannot be non- existent. For existence of information nothing is necessary besides (beyond) the information itself. Indeed, though we principally cannot prove the uniqueness of the "Information" Set, and so formally cannot exclude some external Creator, Who developed "Logos" set, etc., and so created the Set (and thus Who should exist "in a longer time then always"; though that is possible in principle), it seems quite evident, that, even if something External to the Set exists, *then this External cannot be represented as some information*, and so – since now humans evidently observe only some informational patterns/systems, including themselves, and so completely for sure they, and everything else is/ are only the Set's elements, it looks as that this "External Creator problem" isn't now actual; and, besides, it looks is superfluous at least now, most of "Logos" set elements are well rationally formulated in purely Information domain, besides now only the element "Energy", however this Energy problem rather probably will be solved at further development of this concept.

So Matter is completely for sure some informational system/element of the Set, which, as that follows from observations, is based on some set of fundamental and universal laws, links and constants, so Matter's constant evolution is very effectively can be described and analyzed by using mathematics.

The metaphysical phenomenon "Consciousness in Idealism" is purely transcendent concept, which by no means was observed by humans really, however for sure (see below) there exist the phenomenon, and also some the Set's element, "consciousness on Earth", which is rather evidently "immaterial", nonetheless its operation is observed, expressible, and which works evidently only by using information.

Though note, that in the mainstream philosophy and sciences some independent phenomenon "consciousness" mostly doesn't exist, and the phenomenon/notion "consciousness" is defined [Oxford Dictionary] as

1. the state of being able to use your senses and mental powers to understand what is happening I can't remember any more — I must have lost consciousness. She did not regain consciousness and died the next day.

2. the state of being aware of something synonym awareness his consciousness of the challenge facing him class-consciousness (= consciousness of different classes in society)

3. the ideas and opinions of a person or group her newly-developed political consciousness issues affecting the popular consciousness of the time.

- etc., i.e. first of all "consciousness" is the state of a human. However, in this case evidently the questions arise – what is this "human", and why this "human" is as (s)he is? including why can be in this "state"?, which principally cannot be answered in the mainstream, where "consciousness on Earth" – and so "human" are equally really transcendent as some "officially fundamental" "Idea", "Spirit", etc.

Really eventually in the mainstream philosophy and sciences everything, not only "human", is/are transcendent, and is cognizable eventually only on some instinctive level; including at studying of everything in Matter, where really "particles", "bodies", fields, etc., are some transcendent objects that exist and interact for some transcendent mystic reason and by some transcendent way so as they exist and interact. Correspondingly humans perceive the environment, i.e. what exists and happens in Matter, themselves, and societies eventually mostly instinctively, while again the problem – what is this instinctive level, and why it is such that despite of the transcendence above humans are really able to use "senses and mental powers" to understand something, etc.? – remains be transcendent.

Let us consider these fundamental phenomena/notions more specifically.

## 7.1 Matter

### 7.1.1 Logical base of Matter and Matter's spacetime

So, Matter is a set/system of some elements – elementary particles, which mutually interact using mediators of at least 4 known (in mainstream physics of 3 known) now fundamental Nature forces (fields), some systems of the particles and the fields, etc., where all elements interact using exclusively true, complete, and completely rigorous information. I.e. Matter is somewhat similar to the computer. The premise that Matter is some logically organized system isn't, of course, new. It is enough to recall, e.g., Pythagoras's "All from number" and Plato's "All from triangles" doctrines. A number of specific hypotheses that our Universe is a large computer appeared practically at once with the appearance of usual computers (see, e.g. [Zuse, *1969*]; [Penrose, 1971]; [Fredkin, Toffoli, 1982]; [Tegmark, 1998]; [Lloyd, 1999]; [Schmidhuber, 2000]; [Lloyd, 2002]; [Margolus, 2003]; [Gershenson, 2007]; [Tegmark, 2007]; [McCabe, 2008]; though this list can be much longer). An assumption that Matter (Universe) is some set ("ensemble") exists at least since 1998 [Tegmark, 1998].

In philosophy corresponding concept is known as "Informational Realism" [Floridi, 2004]:

"…Informational realism (IR) is a version of structural realism. As a form of realism, it is committed to the existence of a mind-independent reality… it is suggested that an ontology of structural objects for OSR (*ontic* structural realism ) can reasonably be developed in terms of informational objects… outcome is *informational realism*, the view that the world is the totality of informational objects dynamically interacting with each other".

However, all these suggestions are nothing more than hypotheses, surmises, based, first of all, on the remarkable adequacy of languages, especially mathematical, to external reality. Including the suggestion that the information is a base of Matter is Wheeler's *"it from bit" doctrine* [Wheeler, 1990]:

> "…It is not unreasonable to imagine that information sits at the core of physics, just as it sits at the core of a computer. It from bit. Otherwise put, every 'it'—every particle, every field of force, even the space-time continuum itself—derives its function, its meaning, its very existence entirely— even if in some contexts indirectly—from the apparatus-elicited answers to yes-or-no questions, binary choices, bits. 'It from bit' symbolizes the idea that every item of the physical world has at bottom—a very deep bottom, in most instances—an immaterial source and explanation; that which we call reality arises in the last analysis from the posing of yes-no questions and the registering of equipment-evoked responses; in short, that all things physical are information-theoretic in origin and that this is a participatory universe."

- really has no real rational grounds, and is, though reasonable, purely ad hoc assertion, from which – and from any other hypotheses above – really in physics no some essential implications followed.

Except, however, two really outstanding findings in XX century, which, though were really also transcendent, were brilliant guesses, but in this "Information as Absolute" concept *become to be quite natural.* The first one is the C. F. von Weizsäcker's 1950-54s [Weizsäcker, 1952, 1955] idea of the quantum theory as of a theory of binary alternatives ("UR-hypothesis"), where von Weizsäcker to underline the fundamentality of the alternatives (bits) called them "Urs" – the alternatives in German mythology that are foundation of World.

The hypothesis has rather weighty reasoning. Von Weizsäcker ([Lyre, 2003])

> "…Mathematically, … had just stumbled…" about a well-known fact that any vector in 3-D space can be represented also by some combination of two-dimensional spinors, from what follow at least two important consequences: (i) – three-dimensionality of the "position space" (i.e. the space here), and (ii) - any object which in quantum theory is represented by a Hilbert space can be described in a state space which is isomorphic to a subspace of tensor products of two dimensional complex spaces…"

The Ur-hypothesis on one hand really was essentially rationally grounded – the observable Matter's space is indeed 3-dimensional, and on the other hand the hypothesis solved the one of really utmost fundamental physical problems in philosophy and physics – *why Matter's space is 3D*?

The other outstanding breakthrough in physics was the E. Fredkin, and T. Toffli finding [Fredkin and Toffli, 1982]; [Fredkin 2000]; [Petri, 1967]; [Toffoli, 1980]; [Margolus, 2003], who showed that if some patterns in a system are based on a reversible logic, the system changes at interactions in it without energy dissipation outside the system. In such case, in Matter would not be energy dissipation somewhere in the Set; thus seems thrifty Matter's Creator used this fact; and so in Matter the energy conservation law acts.

Correspondingly (see section 6.1. above) the concrete spacetime of the concrete binary reversible informational system "Matter" has 3 "purely space" dimensions, *X,Y,Z.* Since this system is dynamical system, as that follows from experimental data, the spacetime has the "true time" dimension, *t*, which is absolutely universal and common for all dynamical elements of the Set. Further in this paper, as that is also in the whole informational physical model [Shevchenko, Tokarevsky, (2007-2008), (2008-2010a), (2012-2015), 2021], which is based on this concept, for some reason (see below) instead of "*t*" for the time dimension is mostly used metrics "*ct*", *c* is the standard speed of light.

Besides these four dimensions above Matter's spacetime has once more dimension, which is necessary to implement the degree of freedom of the reverse sequences of changes, which are in a sense "non-legitimate" in the time, as some "travels backward in time", what is principally prohibited in the time. This dimension is really a specific space dimension, however it is actualized in many traits in the Matter like the "true" time, *t*. This dimension is called in the concept and the model "coordinate time", "$\tau$", dimension, since that is just the "time what clocks show" [more see below], and mostly further for this dimension the metrics "$c\tau$"is used.

Thus the Matter's spacetime is the [5]4D Euclidian spacetime as an infinite empty container, where Matter exists and constantly, because of the energy conservation law, changes, with the metrics ($c\tau, X, Y, Z, ct$), where "$c\tau$" is the "coordinate time" dimension, "*ct*" is the true time dimension, and *X, Y, Z* are 3 "ordinary" space dimensions. The dimensions (see section 6.1. above) above, are principally infinite by definition of Space and Time; and so, for example, in this spacetime there can be placed and evolve infinite number of "Matters", i.e. informational systems in the Set that are based on the identical logical design as the observable Matter has, and on infinite distances from each other.

Finally, note here also the historically important problem – is the Matter's spacetime absolute or not?

This problem did not exist in philosophy and physics (for a long time, when, in those times "naturephilosophical" Newton's definition of absolute space and time were held as true) till the fundamental Nature EM force was discovered, or even in first years after development of the Maxwell-Lorentz theory, where EM objects, events and processes existed and happened as some disturbances in some "ether", fixed in corresponding absolute Euclidian space. However, in late 1800s it became clear, that seems as the application of very mighty Galileo relativity principle to EM processes and events results in some paradoxical consequences, as, say, the "relativity of simultaneity". It also seemed that because of the principle it is impossible really to observe absolute space and corresponding absolute motion of bodies in 3D space.

H. Poincaré wrote about the absolute motion in "Science and hypothesis" [Poincaré, 2190:

> "… Again, it would be necessary to have an ether in order that so-called absolute movements should not be their displacements with respect to empty space, but with respect to something concrete. Will this ever be accomplished? I don't think so and I shall explain why; and yet, it is not absurd, for others have entertained this view… I think that such a hope is illusory; it was none the less interesting to show that a success of this kind would, in certain sense, open to us a new world…"

However, from that the absolute space even indeed cannot be observed evidently does not follow that it doesn't exist. Nonetheless that was postulated in the first version of the special relativity theory (SR) in 1905 [Einstein, 1905]. It was also postulated that there is no corresponding ("luminiferous") ether, which would be placed in the absolute space, and be a base of some absolute reference frame. So the SR was – and is till now - based on one more postulate that all/every inertial reference frames are absolutely completely equivalent and legitimate.

From the last postulate any number of evidently meaningless physical, logical, biological, etc., consequences directly and unambiguously follow, the simplest one is the well known “Dingle objection to the SR” [Dingle, 1967] and its more known and more complex version “twin paradox” [Shevchenko, Tokarevsky, 2018], etc.

From even one meaningless consequence, which directly and unambiguously follows from the postulates above, it completely rigorously follows by “proof by contradiction” that Matter’s spacetime is absolute; and that follows from the definitions of Space and Time in Sec. 6.1 above as well. However, these SR postulates have been stated as true postulates in physics till now.

Correspondingly observation of the absolute motion, i.e. the motion of a body in the absolute 3D space, is only a technical task, which can be principally solved, as that is shown in the informational physical model, and the absolute velocity of a pair of clocks can be measured yet now [Shevchenko, Tokarevsky, 2016].

#### 7.1.2. Actualization of the logical base — Matter as “computer”

The hypotheses that Matter is some analogue of a computer, in which some program code always operates were, and are, rather popular in philosophy and physics, see, besides references above e.g. [Lloyd, 1999]; [Fredkin, 2000]; [Schmidhuber, 2000]; [Lloyd, 2001]; [Margolus, 2003]. This follows from the fact that (fundamental) Nature laws are comparatively simple, the number of the laws is not large; at that, the laws (as well as the elementary particles, or more correctly their taxonomy, which is relevant to the particles’ structure) can be reduced to a number of the groups of high-level symmetry.

To build a computer, as is well known, some simplest controlled logical elements allowing realizing main logical operations in the computer are necessary. So it is plausible to suggest [Shevchenko, Tokarevsky, 2007] that the computer “Matter” is built on a base of such elements, that are in a way analogues of Weizsäcker’s “Urs”, and constitute a dense lattice in the spacetime, some analogues of Penrose’s “spin-network units” [Penrose, 1971], and “causal set” [Sorkin, 1991], “Space-time points in causal space” [Finkelstein, 1969], etc., which, nonetheless, principally differ from the analogues above. In the informational physical model and here that are called “binary reversible fundamental logical elements” (FLE).

The FLEs themselves are naturally some informational structures also. Since in the “Information” Set every of Its elements is always connected with all other ones by some informational relations, to make up some stable structures from the FLEs, the FLE might have the property that informational connections inside the FLEs and between FLEs, including dynamic ones, in the informational structure “Matter” must be much stronger then any other FLE connections in the Set.

A human does not observe structures and work of the FLEs directly, similarly, as he, e.g., doesn’t observe flipping of elementary logical gates in a PC, and only sees the results of the flipping, say, pictures on the display. Nonetheless, he sees (measures by the instruments) some results of the work of “program shell” that governs objects/events/effects/processes in Matter, which are eventually some FLE-structures as well.

As in the case of ordinary computer, for the FLE it is sufficient to have, at the minimum, two possible states (“0” and “1”), i.e. to have a possibility to form 1 bit of information, and to have some control inputs to change a FLE state – “to flip” the FLE by an external signal. So simplest cause-effect (dynamic) operation in Matter is the flipping of a FLE that is carried out during the minimal time interval $\tau_0$. If we assume, also, that the minimal length in Matter is the "size" of the FLE $l_0$, then maximum speed of propagation of an information in Matter is $c = l_0 / \tau_0$.

In this concept – and the informational physical model – it is quite rationally premised, as that rather reasonably follows from existent experimental data, that minimal intervals $\tau_0$ and $l_0$ are Planck time $t_P$ and Planck length $l_P$; correspondingly maximum speed of propagation of information is equal to speed of light $c = \frac{l_P}{t_P}$.

Really FLEs in Matter aren't some direct analogues of computers' logical gates, they have (1+3) independent degreases of freedom at changing of their state "flips", and compose the Matter's "aether" – the *[5]4D dense FLE-lattice,* which is placed in the [5]4D Matter's spacetime above, the space dimensions, of which, i.e. the *cτ-, X-, Y-,* and *Z*-dimensions correspond to 4 degreases of freedom at changes of FLE states, which (the lattice) is ultimate base of Matter and *everything in Matter is/are some disturbances in the lattice.*

Besides the 4 utmost fundamental "kinematical" operation abilities of this "FLE-hardware" above FLEs have at least 4 logical "marks", which correspond to four known now logical constructions "fundamental Nature forces" – "Gravity", "Weak", "Electric" and "Strong" Forces, which are actualized in the logical systems "particles" by a specific way, so particles compose atoms, molecules, and further the innumerous diverse material objects.

Returning once more to the Matter's spacetime problem note that in mainstream philosophy and physics it is stated that Matter's real spacetime is the postulated in the special and general relativity theories, standard now in physics, and in most versions of ontology of "space" and "time" in philosophy, the pseudo-Euclidian 4D Minkowski space (special relativity) and 4D pseudo-Riemannian space (in general relativity), for which a number of "relativistic properties and effects" in these theories are postulated,

- that are actualized at some impacts on these spaces by material objects – when a system of material bodies, i.e. of scaled rules and synchronized clocks, "inertial reference frame" moves, it "contracts space" and "dilates time", and, in turn, these "contracted" space and "dilated" time really contract length of bodies, slow down clock tick rates, etc. In general relativity some real interactions in system "masses-spacetime-mass" are postulated, when masses "curve" the spacetime, and this "curved" spacetime forces material bodies to move "along geodesics".

Really, again, the Matter's spacetime is the infinite logical [5]4D "empty container", the dimensions of which are determined by logical construction of FLE, which is ultimately strong, and so in Matter nothing exist, what could to change by some way any FLE and to impact/transform by this way the spacetime. Really the postulated in special relativity properties and effects are the postulated in the theory illusory interpretations of existent experimental data, first of all as that in the Lorentz transformations the space and time coordinates/variables relate to every point in the Matter's spacetime (Minkowski space), what fundamentally is wrong, the variables really relate only to the spacetime points that are occupied by material bodies, which do really are contracted at motion in 3D space, and, at that, internal processes in moving bodies do are slowed down. More about what are the Lorentz transformations, etc., see [Shevchenko, Tokarevsky, 2021, 2022]

### 7.1.3 Main constituents of Matter – particles and fields

Matter is the system of innumerous diverse material objects –gases, liquids, bodies, etc., cosmological objects, which, nonetheless, all are constituted from a few *particles*, which interact with each other by at least four *fundamental Nature forces*, which act as that some particle that has the concrete "charge" of a concrete Force, e.g. "electric charge" for Electric Force, gravitational charge "gravitational mass" for Gravity Force, which [the charge] radiates this Force mediators that are specific disturbances in the [5]4D FLE lattice, which propagate in the FLE-lattice, and, if hit in some other particle that has the same Force's charge, to this particle some momentum is transmitted. If the charges are large, as that happens on the macro scale bodies that are composed from extremely large numbers of particles, that is observed by humans that (since the other bodies' particles radiate also the mediators that hit into the first ones above) on the bodies some "Newton's" forces impact. Flows of the Forces' mediators are observed on macro scale as the "*Forces' fields*".

As that rather scientifically rationally, and in complete accordance with all existent really reliable experimental data, is assumed – and postulated – in the informational physical model all particles in Matter are constructed and made by the same logical scheme: that are specific disturbances in FLE lattice that are created if some the lattice's FLE is impacted by something with transmission to the FLE some 4D momentum, $\vec{P}$.

If the momentum is practically infinitesimal, this FLE "flips", and causes the flipping of next FLE, etc., so in the (1+3)D space with metrics (*cτ, X, Y, Z*), the sequential flipping appears along the straight line in direction of the momentum, where the "flip-point" moves in the lattice with the 4D speed of light. However, if the momentum isn't infinitesimal, since the flipping cannot propagate in the lattice with a speed that is larger than the speed of light, such impact causes FLE precessing, and so the flip-point propagates along some 4D "helix", what results in that the flipping FLEs compose some close-loop algorithm, which cyclically run with a frequency $\omega$;

- and this flip-point that has the creating momentum, $\vec{P}$ and energy $E = Pc = \hbar\omega = mc^2, m$ is the inertial mass, is some particle that is the "particle as a point", when the 3D projection of the 4D "helix" is observed in 3D space as the "particle as a wave", in QM "wave-particle duality" .

The scheme above is some first approximation scheme, and real motion of particles is more complicated for a number of reasons, for example because of in 4D space some 4D "helix" hasn't its 4D axis vector, that is mathematically impossible, and so the "helix" is some moving tensor, etc., however it well clarifies the "what is particle" problem.

Note also that

(i) — the "radius" of the "helix" is equal to the particle's Compton length $\lambda = \frac{\hbar}{mc}$, the corresponding "helical" angular momentum of the particle's "FLE flipping point" is equal so to the Planck constant $\hbar$, and the Compton length of a particle so is the "spatial" length of the particle's algorithm, while the number $N = \lambda / l_P$ is "logical length", i.e. the number of FLEs in the algorithm. The logical lengths of real particles are large, say, electron's algorithm consists of $N \approx 3.7\text{x}10^{21}$ FLEs;

(ii) — since the $c\tau$-dimension fundamentally differs from the standard space dimensions, particles that are created by the momentums that are directed along the cτ-axis (in the physical model "T-particles") essentially differ from particles that are created by the momentums that are directed along some 3D standard space direction ("S-particles").

Both types particles so move in the [5]4D FLE lattice and in the 4D sub-spacetime with metrics ($c\tau$, $X$, $Y$, $Z$) having 4D velocities, which have identical absolute values being equal to the speed of light, $\vec{c}$, and, in parallel, with the speed of light in the $ct$-dimension.

So S-particles, now we know really only S-particles "photons", always move in the 3D space with the speed of light, while T-particles move in the cτ-dimension with the speed of light only if are at rest in the standard 3D space; and so, if are impacted by some 3D space directed momentums, T-particles reveal inertial "rest mass". S-particles fundamentally obligatorily have inertial masses as well, and, since cannot be at rest in the 3D space, haven't "rest masses", so in mainstream physics publications often it is claimed that photons are "massless", despite of that is principally wrong, all/every particles in Matter fundamentally have some inertia and inertial masses.

Besides the above here exists once more fundamental difference between T- and S- particles. The $c\tau$-dimension is intended for actualization of reversibility of logical operations in Matter, i.e. of the FLE reversibility degrease of freedom at their flips, and so T-particles that are created by directed along positive cτ-axis direction momentums and T-particles that are created by directed along negative $c\tau$-axis direction momentums, are logically different – the first particles are some close-loop algorithms with some command order, the other particles are the same algorithms, but with reverse command orders,

- what is observed at experiments as "particles" and "antiparticles", which, if meet, annihilate.

***Other, besides the particles, FLE-lattice disturbances*** are "*fields*", which were introduced in physics as "they say that in some point in space some Force's field exists, if on a test body that has a charge of the Force and is placed in this point some force acts", at that the first forces were observed in gravitational and electric fields, i.e. actions of two fundamental Nature Gravity and Electric forces. In middle of 20-th century it was experimentally discovered that there exist also two other Forces that are extremely short range Forces that act only on micro scale – "Strong" Force that provides existence of atoms' nuclei, and further of hadrons, and "Weak" Force that causes some type of particles decays. Though at that Gravity was claimed as that it isn't some fundamental Nature force, because in physics standard theory of gravitational objects/events/processes is the general relativity theory, where the postulated in GR as real interactions in systems "mass-spacetime-mass", i.e. when the spacetime is real material force; and so fundamentally differs from all other Forces, which act in the spacetime principally without any affecting the spacetime and reverse, that isn't essential in this case, Gravity is a Force.

In parallel with experimental study of Matter in those times the corresponding theories of the Forces interactions was developed, which describe and analyze the objects/events/processes in Matter, first of all particles and particles interactions on the level, where applications of the "classical" quantum mechanics, which was developed mostly as clarification of how the Electric Force acts on the micro-scale, become be non-effective – the "quantum fields theories". The first such theory was the quantum electrodynamics (QED), further the quantum chromodynamics was developed, essentially basing on the approaches and techniques in QED, which addresses to Strong Force phenomena, and some techniques that address to the Weak Force; these theories recently are united in the "quantum field theory", QFT.

Since QFT addresses to phenomena in Matter on really fundamental level, the fact that Matter – and so everything in Matter - is fundamentally transcendent reveals itself in that QFTs are based on practically completely transcendent postulates, which drastically differ QFTs from other physics, which is based, though eventually also inevitably transcendent, however more directly and rationally grounded on reliable experimental data, postulates, first of all in QFT such physical objects as "virtual particles", as the real mediators of all Forces and in some cases virtual "ordinary" (not mediators) particles, are postulated, and, besides, the main real QFT tenets are as [Tong, 2022]

> "…. QFT is the most successful scientific theory of all time …. The big appreciation, then, is that it's the fields that are really fundamental, that the electric and magnetic field is at the basis of everything. And little ripples of the electric and magnetic field get turned into little bundles of energy that we then call photons due to the effects of quantum mechanic. And the wonderful big step, one of the great unifying steps in, in the history of physics, is to understand that that same story holds for all other particles. That the things we call electrons and the things we call quarks are not themselves the fundamental objects. Instead, there is spread throughout the entire universe something called an electron field, exactly like the electric and magnetic fields. And the particles that we call electrons are little ripples of this electron field. And the same is true for any other particle you care to mention. There's a quark field — in fact, there are six different quark fields throughout the universe. There are neutrino fields, there are fields for gluons and W bosons. And whenever we discover a new particle, the most recent being the Higgs boson, we know that associated to that is a field which underlies it, and the particles are just ripples of the field.…"

I.e. in QFT "fields" obtain drastically other sense than that is in classical theories and "classical" QM, where really existent particles create the Forces' "forces fields", which act to particles – QFT fields are some "sources" of particles that appear at the fields excitations, and so, including, there are so many these fields that are spread throughout the entire universe space how many particles exist (now a few hundreds ones are in the discovered particles zoo), and these "fields" have practically no relation to the observed fundamental Nature forces.

All that looks as rather transcendent picture, whereas it looks as completely rational to assume that real Matter's objects, including particles, aren't "virtual", but are real, that interactions by Forces are mediated by real mediators, etc., that inertial masses of particles are some actualizations of the absolutely fundamental phenomenon "Inertia", and so for existence of particles' masses fundamentally there is no necessity in some special – in QFT Higgs – field, etc.,

- and, besides, it looks in this case as rather natural that in QFT principally additionally some evidently transcendent ad hoc mathematical tricks, first of all some "renormalization" techniques, are applied.

So the P. Dirac's words [ Dirac, 1975] about the first QFT theory – QED

> "…I must say that I am very dissatisfied with the situation… because this so-called 'good theory' does involve neglecting infinities which appear in its equations, neglecting them in an arbitrary way. This is just not sensible mathematics. Sensible mathematics involves neglecting a quantity when it is small – not neglecting it just because it is infinitely great and you do not want it!..."

- look as rather rational and essentially applicable to whole QFT now.

In framework of this conception, where Matter – and so everything in Matter ceases to be transcendent, what allowed to develop really rational physical model, including of particles (see above), and two – Gravity and Electric – Forces models, where the Forces are mediated by corresponding real mediators, the mediators aren't particles, however compose the classical gravitational and electric real fields, and where all, particles, just which are the fundamental objects that radiate mediators, and the mediators, are completely real disturbances in the FLE-lattice, etc., more see [Shevchenko, Tokarevsky, 2021]

Though in the model the QFT postulate above about always and everywhere existent "electron", "quark", etc., fields, becomes to be formally rather correct – every of the lattice 's FLE has all Forces' marks, and, if is impacted by some concrete way in accordance with some concrete combination of Matter's laws/links/constant that are actualized at concrete interactions, in any space point any concrete real particles can be created. However, again, that really is only a formal resemblance, which, though, can be non-accidental in this case. Though QFTs are based on numbers of principally ad hoc postulates, nonetheless these postulates aren't arbitrary, they are introduced in the theories aimed at fitting the theories with experiments, and applications of the theories in practice, at least QED, so are in some cases well adequate to the reality. Correspondingly the development of really non-transcendent theories will be essentially basing on existent QFTs also, including as answering to questions – why some ad hoc and really strange transcendent postulate has, nonetheless, some physical sense.

More about what happens in Matter, including about what are a few dozens of fundamental problems, and how these problems rather probably should be solving, in physics, including re-formulation of classical and quantum physics in accordance with the scientific definitions of the fundamental physical phenomena above,

- say, from the scientific definition of "time" it follows re-formulation of QM, where, because of that in QM till now really the fundamentally incorrect Newton's definition, with really also fundamentally incorrect "relativistic corrections", the observable "time", *t*, is used, hasn't corresponding conjugate momentum operator $i\hbar\frac{d}{dt}$ ,whereas this operator in QM is postulated as the energy operator,

- and so, say, at developing his relativistic QM equation for electron Dirac introduced completely transcendent "Dirac sea" of "negative energies" [Dirac, 1934], while some "negative energies" fundamentally don't exist in Matter, in QED purely ad hoc the transcendent "Feynman–Stueckelberg interpretation" [Stueckelberg, 1941], [Feynman, 1949] is postulated, that antiparticles move in negative time direction, whereas in Matter, as that correctly is in mainstream physics as well, fundamentally cannot be any back in time motion, and so these, indeed brilliant, guesses, were made only as some violations of existent physics

Really Feynman–Stueckelberg interpretation is really adequate to the reality since antiparticles really move in negative cτ-dimension direction since are created by momentums that are directed in negative cτ-dimension direction; whereas in the standard now in physics 4D Minkowski space the time coordinate is some mix of the true and coordinate times. Operator $i\hbar\frac{d}{dt}$ , more correctly $i\hbar\frac{d}{d(ct)}$ isn't really an energy operator, that is, quite naturally, conjugate to *t* operator of whole 4D momentum, as that are corresponding momentums operators $-i\hbar\frac{\partial}{\partial x_i}; i = 1,2,3$, as the conjugate operators to 3 space observables *x, y, z*, etc.

- more see whole informational physical model [[Shevchenko, Tokarevsky, 2021, 2022], here only a couple of additionally notes: really not only QM and QFT should be re-formulated, but the "classical" theories also; and, the informational model is rather essentially testable, including predicted in the model quantum nature of Gravity Force rather probably can be observed, at least in experiments with photons yet now [Shevchenko, Tokarevsky, 2007-2008, 2011]

## 7.2. Consciousness and AI

Consciousness, as the utmost basic phenomenon/notion in the mainstream philosophical doctrine "Idealism" (where also other terms are used, e.g., "Idea", "Spirit", etc.), in contrast to observable "Matter" in Materialism, is completely transcendent non-observable phenomenon, and so Idealism practically doesn't differ from any developed religion, practically unique difference is in terminology, though even that isn't completely correct; say, Hegel's Idealism texts and Veda and Upanishads texts sometimes look as rather similar, while really omnipotent and omniscient in Idealism "Ideas", "Spirits", etc., really don't differ from, say, God in Abrahamic religions.

So here we don't consider ontology of the philosophical phenomenon above, and address to the really observable on Earth informational system "consciousness on Earth", which, as that was pointed already here above, doesn't exist in mainstream philosophy and science, first of all biosciences and physics, as some independent phenomenon and is defined in the mainstream as a state of human when (s)he is "conscious". Though this view looks as so questionable, that a large, though non-authoritative, part of philosophers either following some religions, or independently, admitted that something that is non-material exists in humans. An example is the Descartes's the *res extensa* and the *Res cogitans realms* and *"I think, therefore I am";* in recent mainstream philosophy the phenomenon/notion "mind" exists, and problems as "hard" and "easy" "body and brain" problems are actively debated in tens of years' discussions, etc.,

- and all that happens without any rationally grounded results, because of that really in the mainstream everything in this case, i.e. phenomena/notions "I", "think", "body", "brain", "mind", etc., are transcendent, and though all these notions are applied in everyday and scientific humans' practice in many cases rationally and adequately to the objective reality, that happens only because of these terms – and everything else – are defined and used by humans only as some instinctive actualization of some instinctive sub-conscious a priory information that is, first of all "automatically" "written" in some human's structures, more see below.

Note, though, that albeit the situation above in the mainstream exists because the fundamental transcendence of everything in the mainstream, however even in this case it looks as rather strange, since existence of something, that practically for sure is non-material in humans looks as evident, as that, e.g. was pointed by Granville Sewell (Sewell, 2022)

> "…Peter Urone, in his physics text College Physics, writes, "One of the most remarkable simplifications in physics is that only four distinct forces account for all known phenomena."
>
> This is what you have to believe to not believe in intelligent design: that the origin and evolution of life, and the evolution of human consciousness and intelligence, are due entirely to a few unintelligent forces of physics. Thus you must believe that a few unintelligent forces

> of physics alone could have rearranged the fundamental particles of physics into computers and science texts and jet airplanes and nuclear power plants and Apple iPhones. ...."

- i.e. it is quite evident that though humans cannot, say, to ground rationally the state that material structures cannot think as that humans do, however humans evidently made something that Matter doesn't make for sure, and so at least jet airplanes, nuclear power plants and Apple iPhones are designed and made by something,

- and, since humans, as their bodies and brains, are evidently some compositions of completely material atoms, molecules, etc., this something so is fundamentally non-material, nonetheless rather evidently resides in humans, and governs the humans' behavior.

Including from the above it follows that not atoms, molecules, etc., in humans, but just this something has the state "being consciousness", etc., and thus this, evidently essentially independent on body and brain, something is some fundamentally *non-material* informational system "*consciousness*".

In the concept both, "Matter" and any "Consciousness" [from existence of the observable consciousness it looks as quite rational to assume that there can be other consciousnesses in the Set, which are some other versions of specific system "Consciousness", as the "consciousness on Earth" is as well], are utmost commonly scientifically defined — "Matter" and "Consciousness" absolutely for sure are nothing else than some informational systems – the Set's elements, so are made from the same stuff "Information", and in accordance with the same "Logos" set's elements; whereas "Information" (and yet now most of "Logos" elements as well aren't) isn't transcendent, and it, and so any informational structure as well, can be principally rationally cognizable, (what is "cognizable"? – see below).

Note here also, that when we considered above "Matter as "computer"" really that was in certain sense superfluous — from the above it follows that any informational system of elements is always something like "computer+program shell" system, where "hardware" is the elements, and "program shell" is the concrete the system's basic set of laws/links/constants, in accordance with the elements interact, composing just this system, exchanging at that by some informational messages, which use concrete language that is understandable by elements in the system — as that happens in Matter.

Including "Matter" and "Consciousness" are some systems, which, however, have fundamentally different basic sets of the laws/links/constants, and so are fundamentally different; the main difference is in that Matter *is logically rigorously closed in the Set system*, which so practically doesn't interact with other the Set's elements and thus is essentially stable system that exists at least soon 14 billions of years as the same, i.e. based on the same set of fundamental laws/links/constants, system;

- whereas any *Consciousness* is fundamentally *open in the Set system*, and, as that is one of the utmost specific properties of just any consciousness in the Set, is that consciousness principally is able to obtain, and *instinctively – "automatically", as that Matter's elements do, applying the Logos element "Logical Rules*", logically analyze, any arbitrary information in the "Information" Set. At that, however, because of any consciousness has fundamentally limited capabilities at obtaining and processing of the principally infinite in this case information, every result of the processing is always at least partially uncertain, and sometimes erroneous and illusory;

- and, at that, if a consciousness obtains some information, about which she hasn't some previous information, the consciousness assigns to this information the label "this information is non-understandable", and further, if that is necessary, or that is interesting ("*Curiosity*" is some, just utmost specific property/resident *utility* in the "program shell" of any consciousness), studies the Set's element, from which this information is obtained – again "no understanding" state is just fundamental specific state of any consciousness.

All that in consciousness is in principal contrast to what happens in Matter, where every of Matter's elements, i.e. particles, bodies, fields, cosmological objects, always completely knows all Matter's laws/links/constants, including, e, g., whereas in program shells of any/every system of elements, including in both, Matter and consciousness, in every element logically obligatorily resident *utility* "*Self-awareness*" exists and always runs, so every particle, and any other material object has complete "Self-awareness"; say, every electron completely knows that it is just this electron and principally nothing else, etc., so at interactions uses/exchanges by only true information, and behaves after obtaining some concrete message again in complete accordance with the basic Matter set above. Or, by another words, since the Matter's basic set "is written" in every Matter's element, Matter isn't some "whole" computer, it is an automaton,

- whereas in any consciousness, since, again, she is fundamentally open in the Set system that so exists and operates in principally so at least partially uncertain environment, processing of any information is always also at least partially uncertain, including that happens when the resident utility "Self-awareness" runs.

Besides the pointed above specific ("Curiosity"), and fundamentally common, but specifically running, "Self-awareness", in any consciousness a number of other utilities run, first of all "*Providing of self-stability*" and "*Seeking for the self-development*", which, of course – and that clearly follows from observations, *are principally absent in Matter's program shell.*

Note that the specific consciousness utilities above really run first of all aimed at providing the consciousness stability, and really be governed by the utility "Providing of self-stability" – to provide stability in the arbitrary, and so sometimes unstable, hazardous and even aggressive and dangerous, environment in the Set it is necessary constantly to study the environment (to be constantly curios) and to develop functional possibilities to block, prevent, etc., any possible destructive situations, for what is necessary to develop more and more functional possibilities for obtaining and analysis of information.

These basic utilities in concrete Consciousness versions run rather specifically, including in "consciousness on Earth", which specifically provided self-stability first of all by that these consciousnesses reside on the material, and so essentially stable, residences, by which, besides, they obtain energy that is stored in Matter, which [energy] is fundamentally necessary for consciousness to be some dynamical system. So in this consciousness, including, of course in the "homo sapiens sapiens consciousness" version, in the program shell corresponding specific utilities that govern the consciousness herself, and further the material residence behavior, aimed at providing stability of the residence, i.e. utilities that compose a set of *"material interest" utilities* – first of all providing safety, comfort, food, reproduction, etc. of the practically material bodies.

Thus any Consciousness version is a "whole" computer, i.e. her hardware and program shell contain the main functional modules, i.e. "BIOS", "processor",

“random access memory”, and some specific utilities that organize work of the whole consciousness’s set of functional modules and structures, which [main modules] practically for sure exist and work mostly outside Matter and Matter’s spacetime, i.e. in an absolutely fundamentally existent space about which humans know nothing, since the practically for sure existent some “elementary logical gates” in consciousness, the degreases of freedom of which “determine” consciousness space, as that FLEs in Matter are, are completely unknown; and so, say, if in Matter it is possible to establish some etalon for measurement of bodies’ lengths and distances, say, “meter”, but nobody now knows, say – how many meters long a human’s thought is in some unknown consciousness space dimension, etc.

Nonetheless essential part of the consciousness modules is “materially written” in some specific bodies’ structures, first of all in the brain and in nervous systems, and so “consciousness on Earth” operates also in Matter’s space, when she governs, using some unknown “semi-material” forces, practically material living beings’ organisms, including systems “body+brain”. Both, Matter’s and the consciousness’s, spacetimes share the one true time dimension which is fundamentally obligatorily common for all dynamical patterns/systems in whole Set.

Correspondingly the information about environment firstly is detected by practically completely material body’s sensors – eyes, ears, etc., when in some practically material neurons some completely material electric pulses appear, further these pulses are transmitted and filtered by a chain of some lesser material structures, etc. — up to the functional modules and structures that decode the information into language of the non-material modules, where the final processing happens. At that a wide spectrum of certainty is used – from, say, rather certain eyes picture of visible environment up to essentially common signals “feelings”, for example “feeling of pain”, and “emotions”.

So seems ~99% of input information is processed by some standard algorithms, either as instincts, when some input signals are processed by some typical ways, or at the internal transformations of the initial information above, on some “sub-conscious” levels and modes of operation; and only ~1% is processed “*consciously*”, i.e. in the “*mind mode of operation*”, which in mainstream philosophy and science is known/defined as that “humans have mind”, something like happens in computers, where ~99% of information is processed in computer functional modules, and only results of the processing are downloaded on monitor. Note here, though, that just in “mind mode” rather arbitrary illusory interpretations of sub-conscious informational processing results happen, including, say, just in mind mode in many human consciousnesses’, first of all in sciences, the utility “Self-awareness” runs in a strange way – they think that don’t exist, and only are some “states of someones”.

More concretely about the “consciousness on Earth” see the first approximation functional consciousness model [Shevchenko, Tokarevsky, 2018a], finally here touch a few “technical” points.

The first one relates to one of main mainstream philosophical branches “epistemology” — in this concept we have answers to the *really main epistemological – and practically so important in any science – questions “so what studies of what?”, and “why the first what sometimes adequately, and sometimes illusorily, to the objective reality studies the second what?*”

Both these questions principally cannot be answered in framework of the mainstream philosophy, including epistemology, since in the mainstream both the “whats” in the questions above are fundamentally transcendent, however in this concept the answers are natural: in spite of that consciousness, including the “homo sapiens sapiens” version, and Matter, are fundamentally different systems, however, since both are made from nothing besides only one stuff – “Information”, and absolutely obligatorily in accordance with the same “Logos” set,

- there is nothing surprising in that one informational system, which is able to obtain from, and logically analyze information about, other informational systems, makes that correctly, and sometimes incorrectly – that principally by no means differs, say, from the case when a human decodes information that was created by other human, say, when some linguists decode hieroglyphs that were written on some non-existent now languages,

- and in both cases the processing of information happens in consciousness equally completely naturally, because of that fundamentally cannot be by some other way, and so in humans, in all other living beings – as that happens in Matter, though — happens first of all “instinctively”; though at processing in mind mode some problems appear, see above.

The other point relates to the terms/notions “conscious” and its practically synonym “intelligent”, which are used here above, but require, nonetheless, some clarifications. First of all this term evidently relates to some consciousness, however in the brief functional consciousness scheme above it is shown that the whole system “consciousness” operates, nonetheless, mostly “unconsciously”, or “sub-consciously”, where the informational processing proceeds essentially automatically, being determined by corresponding rather rigorous algorithms [we don’t say here about some pathologies, damages, etc.] that are rigidly written in practically material structures in brain and 2-nd signal system, the terms “conscious” and “intelligent” are applicable only when consciousness processes the prepared information in mind mode, i.e. when from the information a number of rather arbitrary implications can follow, and it is necessary to make some choice, to plan additional study of a situation and possible implications, etc.,

- and all that is governed by the resident and specific consciousness utilities above – “Self-awareness”, “Seeking for development”, “Curiosity”, when some typical situations, methods and results of elaborations of problems, etc., are, including, memorized and stored in long term memory – practically for sure in practically material brains, etc.

All that above it is possible now to simulate on computers, where rather simple resident utilities above in a computer’s program shell — “Self-awareness”, etc., are installed and always running in some background mode; if the computer has also a number of inputs with attached sensors that observe the environment, and the program shell contains some program units that analyses sensors’ data etc. with some programmed aim, including using some criteria, aimed at developing optimal versions of solutions of some problems and developing of the program shell (program learning),

- the simulation of event human’s consciousness operation becomes to be practically identical to what is in humans. Such programs now are rather popular, and have specific name “artificial intellect” (AI) programs. Moreover, in last years some such programs well passed the famous “Turing test”, when in dialogs “computer-human” other humans-experts attempt to define who is the human and who is the computer; and, of course, there can be some automatized technical installations, where some smartphones are produced without direct humans’ control – despite of all everything in the techniques is completely material constructions, and so “Matter makes smartphones”.

The last is, of course, only illusion, all techniques are designed and made completely only by some consciousnesses, Matter fundamentally doesn't make such material objects, and the techniques' programs developed completely only by some consciousnesses, Matter cannot do that even with infinitesimal probability.

On another hand some AI computer indeed well imitate human, besides the functional abilities, say, at a Turing test, it also looks well like human – consists of purely material body – hardware, and "consciousness" – the program shell, and so the science fiction stories about some developed computer civilizations that should replace humans' one look as rather reliable scenarios.

Nonetheless really science fiction fundamentally will remain only a science fiction. Any computer fundamentally processes information completely materially, really information is processed by developed by a consciousness program, and any computer is nothing else than some humans' instrument; as, say, when a human ['s consciousness] solves some mathematical equation on a paper sheet by using a pencil, from that by no means it follows that the pencil solves the equation,

- and, what is fundamentally more essential, every eventually purely material structure "computer" as itself fundamentally can obtain information only about what exists and happens in Matter and to process any information only in Matter's space; and fundamentally cannot communicate with other elements in the "Information" Set – in fundamental contrast to any consciousness. Note also here, that, say, some living beings "bacteria", which, as all/every living beings has some "consciousness on Earth" version, well solve the problems "antibiotics" that some by definition conscious living beings pose for them, though have no brains and even single neuron, processing so corresponding information purely outside Matter; and so in this case it looks as rather reasonable to suppose that really bacteria can be quite intelligent.

### 7.3 Life

The problem "what is "Life" and how it appeared on Earth" was quite natural for humans seems yet tens thousands of years ago, and so the questions were answered a long time ago in religions, where the answers are frankly claimed as some transcendental truths that were given some selected humans, including in this case, say in Abrahamic religions all living beings were created by the God.

After the society, including productive forces, developed up to the level when for some people it had become possible to have free time for contemplations, including for attempts to find, instead of transcendent religions' dogmas, some rational answers to the questions the corresponding branch of humans activity – "philosophy" – was formed, that in "classical" Antic Greeks version had three main subjects for study – "Physics", "Logics", and "Ethics", from which a couple of hundreds of years ago a number of nature sciences singled out, which had/have concrete subjects for study — objects/events/effects/processes in humans' environment, first of all in Matter and in Life. About how that was and is at studying Matter see the above, here some brief notes to the "Life problem".

As that was pointed above the two known now fundamental informational systems Matter" and "Consciousness" are fundamentally different. Currently humans know only one Consciousness' version, "the consciousness on Earth", diverse versions of which every living being on Earth, including humans, have. The main differences – and similarities – between a Matter and any Consciousness in the Set are pointed as well – both systems are made in accordance with the same "Logos" elements, both are some "computer+program" systems, where fundamentally similar exchange by, nonetheless, concretely fundamentally different, logically organized information between the systems' elements proceeds;

- however these systems are fundamentally different, since are based on fundamentally different sets of the basic laws/links/constants — and fundamentally different elementary hardware FLEs; and so, though in Matter all/every elements, know physics absolutely completely, what any human never will do, however, at that, all/any material objects/structures fundamentally don't know – and fundamentally aren't able to know – anything else, thus everything in Matter fundamentally isn't, and never can be, "conscious".

Correspondingly every of both, Matter and Consciousness, exists and changes in essentially different spaces, though consciousness on Earth operates also in Matter's space, when she governs, using some unknown forces, practically material living beings' organisms, including systems "body+brain". Both corresponding spacetimes share the one true time dimension which is fundamentally obligatorily common for all dynamical patterns/systems in whole Set (more about what are Matter's space/time/spacetime see [Shevchenko, Tokarevsky 2021] and above.

Thus, there principally cannot be some "emergence" of any consciousness from any material structure, and, besides, from that any, even simplest, living beings, say, bacteria, and cells at all, have logical, functional, and "material" constructions that are well more complex than a smartphone, and so by no means even these practically material structures really could "emerge" purely in Matter as well, and so for sure are created, i.e. designed and made, by some consciousness.

Nonetheless the mainstream philosophy and sciences, first of all in neuroscience and physics, in many existing now "theories" "models", "solutions", etc., principally reject "Creationism", an example [Rennie 2002]: (Creationism is some false concept because of)

> "…Creationist arguments have not stood up under scrutiny. Dozens of scientific and educational organizations reject the general attempts to establish Creationist accounts by misinterpreting or focusing on gaps in scientific knowledge. The National Center for Science Research, for example, writes: "Evolutionists suggest the new information arises from mutations of existing genes or duplicated copies of those genes. Given the complete absence of evidence for their own theory of 'intelligent design' — a theory that has produced not a single scientific paper in a peer-reviewed journal — they instead seek 'confirmation' of their views in controversies about evolutionary biology. "…."

- what is rather evidently scientifically impossible statement. Both, the Darwin's "Origin of Species by Means of Natural Selection" theory and the recent evolutionary biology, really don't tell about just "Origin of Species", they tell about "Origin of Species form already existent Species", and form that after Darwin in the evolutionary biology the "Origin of existent Species" mechanism was really clarified as result of genomes mutations, by no means it follows the origin of species on Earth – in first couple of billions years of Earth existence on Earth there was no any genes.

So really there exists only one really scientifically grounded answer to the question how Life, at least as the first practically material biostructures, was created on Earth – that for sure was done by some conscious being, and further these structures are used by the informational system "consciousness on Earth" (which, including could the "some conscious being" above, some "consciousness out Earth") as her stable residence in the Set, the source of energy, and some stable auxiliary platform, where some functional modules, that enforce this consciousness ability to obtain, and to analyze the obtained, information about the environment and herself, are placed.

In this case two main cases are: in one case there could exist some consciousness in the Set that designed and created both, the first version of consciousness on Earth and her at least first residences, and the case when the informational system "the consciousness on Earth" could, in principle, exist in the Set in parallel with possible Matter's Creator even before Beginning of Matter, and only used the opportunity to make the stable residence when such possibility occurred. Now humans have no rational and confident enough information from which some rational choice between the versions above could be derived.

After that, "the consciousness on Earth" developed the practically material residence in accordance with the principal points and corresponding consciousness utilities above, *what is as evidently observed trend "more and more outside Matter into other Set's regions*", from the first biostructures up to the "homo sapiens sapiens" consciousness version. The last has well developed ability to obtain and to process information in the highest, "mind mode", mode of operation, when information is processed abstractly i.e. in some cases without direct relation to what happens in Matter, or somewhere else in the observed environment.

Finally note here the common for the last sections' two problem. First one is - so why and how did the concrete informational system "Matter" appear in the Set?, and though it is rigorously true that any consciousness fundamentally cannot "emerge" from any material structure, since any/every material structure is some rigorously closed logical system, whereas, say, material structures can be – and are – constructed and created by a consciousness; and so it looks as rather probable that Matter was for some reasons designed and created by some extremely mighty Consciousness in the Set [more see below], however that by no means clarifies the problem – so why/how some consciousness can appear in the Set at all?

In this case it looks as rational to suggest that that can happen, for example, if in the Set some informational systems accidentally appear after some arbitrary, strong enough energetic impacts, and in some such "informational chaotic" systems, which aren't logically rigorously closed, but in which some primitive versions/logical constructions of the consciousness's fundamental utilities "Providing self-stability", and "Seeking self-development" are formed,

- than at least in some cases some of such systems could exist for a long time enough, enforcing their abilities at providing self-stability and abilities to obtain and analyze the information in the Set – and so being more and more stable in the Set; seeking for next and next energy sources for more and more "conscious" operating; and, eventually, when some consciousnesses, at studying of what happens in the Set, have "consciously" understood what the absolutely fundamental phenomenon "Energy is, such consciousnesses become to be able to create rather arbitrary informational patterns/systems in the Set. Rather probably the "consciousness on Earth" is till now in some initial position on this way.

## 8. The problem of Beginning and evolution of Matter

Ad interim let us remind here that from the properties of information it follows that any Set element, including "Zero statement" contains the Set totally in the "Not-I" part; i.e. as the negation, where the information is maximally compressed. (Note, though, that there are a lot of other types of more specified information compression when a fixed information contains in some tacit form possible corresponding information completely. An example: all information that can be obtained in some theory, or more correct, almost all information, if we recall the incompleteness theorems, is contained in the theory's axiom system. All further development and applications of the theory (theorems, tasks, calculations, etc.) do not create any new information including dynamic one in addition to the information that the axioms tacitly contain. L. Wittgenstein wrote: "Proof in logic is merely a mechanical expedient to facilitate the recognition of tautologies in complicated cases." [Wittgenstein, 1921: point 6.1262]. In reality not only proof of something provable [e.g. of theorems] is "a mechanical expedient"; "a mechanical expedient" is yet the formulation of any provable (for given system of axiom) problem, e.g. of a theorem itself);

- and so the fixed true information, in form of "up to Beginning statement" *"there is no this Matter, as well as Its evolution*", existed in the "Information" Set "always", "absolutely long before" the Beginning. And this "Book of Fates" for Matter formally consisting of only one sentence, contained all and absolutely exact data about the Matter, including absolutely complete and exact information about the cause and the method of Creation, as well as about everything what in corresponding time will happen with every element "Matter" in the Set, with every elementary particle and system of particles.

That is Matter was not created "absolutely from nothing" and appeared as some "real" dynamic the Set's element after some – really huge - energy portion has open the conserve can "up to Beginning statement" above.

This "Creation problem" evidently is beyond mainstream science, including cosmology, since in the mainstream Matter and all other, evidently necessary in this case fundamental phenomena, first of all "Space", "Time", "Energy", and really necessary in this case "Consciousness", are fundamentally transcendental phenomena/notions, and so in this case any attempts to put forward any really rational hypothesis, and further to develop some really rational theory/model, logically inevitably can result, and result, when in physics rather numerous attempts happened, only in some transcendent constructions, that are based on some principally transcendent initial premises, even yet in the formulation of corresponding problems, when physics addresses to objects, events, and processes, which humans cannot study now in controlled, or at least observable, conditions.

### 8.1. The "Beginning problem"

This is an utmost fundamental problem in cosmology, and it is rather evidently principally irresolvable in framework of official physics. Physics has no reliable data about the objects, events and processes that could exist, appear, and happen at Beginning. Nonetheless a number of theories exist in physics, and in the standard cosmological "Big Bang" model [Weinberg, 208] in [van de Vis, Sfakianakis, et, al., 2019] it is suggested concretely that

> "…. As the Big Bang theory goes, somewhere around 13.8 billion years ago the universe exploded into being, as an infinitely small, compact fireball of matter that cooled as it expanded, triggering reactions that cooked up the first stars and galaxies, and all the forms of matter that we see (and are) today….."

- in spite of that the existent physics principally is not applicable to this "infinitely small, compact fireball of matter", etc., and so principally isn't able to rationally suggest – which, why and how some reactions cooked up the first stars and galaxies.

As well as to the next steps of Matter's creation, when in the model

"…more explosive phase of the early universe at play: cosmic inflation, which lasted less than a trillionth of a second. During this period, matter — a cold, homogeneous goop — inflated exponentially quickly before processes of the Big Bang took over to more slowly expand and diversify the infant universe..."

- existent physics knows absolutely nothing about what was this "cold, homogeneous goop"; why "it inflated exponentially quickly before next processes of the Big Bang", by what reason this "inflation" stopped; and further by what reason and how that "took over to more slowly expand and diversify the infant universe", etc.

Nonetheless, there exist, basing on existent astrophysical data, a number of seems as rather rational points in standard model of Matter's evolution after Beginning, including, if we do not take into attention the remark above, the rather rational "phenomenological" description of states in Matter evolution above,

- starting from the "space inflation" state/epoch hypothesis [Guth, 1981], [Linde, 2014], when the space, in the standard model for unknown reasons, and by some transcendent way, appeared and exponentially expanded, and that happened at some relaxation of some completely unknown in physics "inflaton" field's singularity, because of "a repulsive gravitational force" (?) [Van de Vis, Sfakianakis, et, al. 2019]. However, the "inflation hypothesis", in spite of these rather questionable points, seems adequately to the reality phenomenologically describes the observed now uniformity of matter density and of the material objects nomenclature on cosmological distances, the nucleosynthesis, etc.

Including the hypothesis in the standard model that during inflation the matter was a cold, homogeneous goop, seems is rather plausible, since that is consistent with cosmological observations. However, that contradicts with the assertion that the matter "exploded into being, as an infinitely small, compact fireball" in this model in the quote above.

The informational approach allows to formulate reasonable physical hypothesis [Shevchenko, Tokarevsky, 2015, 2021], in accordance with the existent experimental data, and with reasonable points in the standard Big Bang model above, such as the inflation epoch, that the Matter after the inflation was rather cold, etc.

In the hypothesis it is suggested that the "Information" Set's element "informational system "Matter"" was created by the other Set's element, "an informational system conscious smart "Creator"", which was indeed extremely smart and could design a logically simple, however functionally extremely complex, effective, and closed in the Set, informational system; and has found in the Set at creation of this system a few huge portions of the till now mysterious, including essentially in this concept, phenomenon "Energy".

Thus – see section 7 above - Matter is based on very nice and smart the simplest binary and reversible logics plus (at least) 4 fundamental logical marks construction, which (marks) humans observe as 4 real fundamental Nature forces, including Gravity, and few universal links and constants, which are "written" in the Matter's utmost fundamental base - in the correspondingly binary [5]4D reversible fundamental logical elements (FLE).

Further this design was actualized into Matter in the next 3 steps – and portions of energy:

On the first step the [5]4D dense lattice of [5]4D FLE was created ("inflation epoch") exponentially, as the result of programmed division, possibly into 2, of possibly one "primary FLE" (as that, say, bacteria spread in a Petri dish, if there are enough resources) in the corresponding Matter's fundamentally infinite, absolute [5]4D spacetime with metrics ($c\tau,X,Y,Z,ct$), Euclidian of course,

- which [the spacetime] "automatically", i.e. by definition of the absolutely fundamental phenomena "Space" and "Time" – see section 6.1. above, appeared at the creation yet of the "primary FLE". Note, though, that this spacetime always existed in the Set, as a sub-spacetime of the Set's whole spacetime. The FLE lattice was cold;

- on the second step, the energy portion with positively $c\tau$-directed momentums was *globally uniformly* pumped in this FLE-lattice, and in the lattice only some primary T-particles were globally uniformly created, what was in fundamental contrast to what is observed at high energy physics experiments now, when fundamentally only pairs "T-particle + T-antiparticle" are created, and so Matter had not antimatter yet at Beginning; what was possible because that the primary particles were completely symmetrical close-loop algorithms, and so creation of only particles was logically permissible,

- and that is the rational solution of one of the oldest fundamental problems in cosmology – the problem "*Why Matter now practically does not contain antimatter*".

. It seems as rather probable that the energy was spent only on the particles creation, and so the "primary T-particles" matter in Matter was probably rather cold again.

However, from existent cosmological data it looks as rationally to assume that the pumping *wasn't uniform locally* – in the lattice [and so in 3D space] some clusters of primary particles were created, where the particles density was radically enhanced, which were some seeds of appearing on next Matter's evolution steps large cosmological objects, first of all – galaxies.

On the third step the primary particles, which in this hypothesis are rather probably Planck mass particles or other simple particles, i.e. that were symmetric algorithms, and have only completely symmetrical gravitational charges, interacted by using only also *completely symmetrical Gravity force*. The result was, rather possibly indeed a soup of existent now, however, because of the angular momentum conservation law, only particles — if the primary particles were the Planck mass particles, then nearly $10^{19}$ "ordinary" baryons were created in an interaction of two particles — which were distributed again globally uniformly – but non-uniformly locally – in the lattice.

In the "soup" unstable particles decayed quickly and – as the standard cosmology asserts rather adequate to the reality – the observable now particles eventually remained, and this soup was rather hot. Hence, cosmic microwave background radiation exists now, however that possibly was not originated at a "singular" temperature, because the energy was mostly spend again on creation of the ordinary particles.

At that, again — quite rationally probable, some the Set's element "conscious smart Designer and Creator" of the logically nice structure above practically for sure did not need to control the step-2 and step-3; and step 1, though. Creator well knew that nothing besides a concrete informational system "Matter" can appear, if a dull energy is pumped in the FLE-lattice; and, say, this Matter could have a number of thousands of galaxies lesser or more, but for Creator that was not essential.

Besides the above the hypothesis presents a rather rational answers on a couple of other fundamental cosmological problems

### 8.2 "What are the "dark matter" and the "dark energy"

The mysterious in physics "***dark matter***" exists in cosmology because of to explain abnormal motion of stars on outer orbits in galaxies - the star move having too large velocities comparing with that should be in accordance with Newton Gravity law, what can be utmost physically reliably explained as that every galaxy contains, besides the visible mostly baryon matter some other matter, which interacts with the visible matter only by Gravity Force, so isn't visible, and is diffusely distributed in space, forming some galaxies haloes; and the mass of the dark matter is in ~ 4-5 times larger than the visible matter mass.

.

In the mainstream physics there exit a number of "dark matter" theories, where, since the corresponding particles that interact only gravitationally are experimentally unknown, and in physics Gravity as some fundamental Nature force doesn't exist at all, because of standard physical theory of Gravity is the general relativity, a lot of, including rather exotic never observable, candidates of the dark matter particles are offered – neutrinos, including non-observed ones, "axions", etc. Really these theories are typical inevitably transcendent constructions, which have, correspondingly, really too indirect relation to the objective reality.

On this hypothesis the really rational dark matter problems obtains rational explanation — it would not be surprising if, say, the "dark matter" indeed exists, being made up from the "primary particles" above. That could happen if during the creation of "ordinary" matter on the 3-rd step only 10-30% of these particles have interacted, and 70-90% of the "relics" exist till now. If these are the Planck mass particles, then the density of the dark matter particles is in ~$10^{19}$ times lesser than the baryons' density, i.e. 3-4 particles in a cube with the size 1000 000 m.

Since the primary particles interact only gravitationally, they interact with "usual" particles at a probability extremely lesser than when that even for neutrinos, and so (i) - the bodies, stars, etc., are practically transparent for these particles, which rotate around centers of some massive bodies, including around galaxies centers, along their single own orbits, forming corresponding haloes, including the whole galaxies' haloes; and (ii) – they are practically non-detectable by humans' instruments, due both to extremely small cross section of interactions with ordinary matter in particles' detectors, and extremely small concentration.

Though if an interaction happens in a detector, that will be well observable, $10^{19}$ BeV is rather observable energy.

Besides it looks as rather rationally to assume, that in the "seeds" clusters there were some local – and small – regions, where the primary particles density was so large, that the primary particles composed compact objects with extreme mass and Gravity field, which have become centers of galaxies, having masses millions, even billions, of stars.

These objects have some interesting physical trait – the strength of created by the objects Gravity field is so large, that escape velocity becomes be equal to the speed of light, and so, say, even photons, if aren't radiated orthogonal to the objects surface, propagate inside corresponding space volume along closed orbits – such objects so practically don't radiate light.

That happens in both existent theories of Gravity – Newton's theory and general relativity, and happens at least provided that the mass, $M$, and radius, $R$, of an such object are in accordance with the equation $R_{gS} = \frac{2GM}{c^2}$, $R_{gS}$ is the radius in GR (Schwarzschild radius) , corresponding radius, $R_{gN}$, in Newton Gravity is two times lesser. At that the radius isn't the object's radius, really it can be lesser than that radiuses above.

The difference of $R_{gS}$ and $R_{gN}$ values isn't principal, however these radiuses principally differ in that when $R_{gN}$ is the radius of some "virtual" surface, which surround some "dark place", the Schwarzschild radius is the radius of the "event horizon" in GR, where solutions of the GR equations become to be singular, and so the event horizon is the border of a "hole in spacetime" – a "black hole" (BH), and so nothing principally can escape from this hole.

Really on the event horizon no singularity exists, the potential and strength of Gravity field increase rather smoothly with decreasing of the distance to the center of the object. So, say, the super massive black holes (SMBH) in centers of galaxies, which, rather probably, are offspring of the "seed" objects that were growing absorbing gas and other matter around at galaxies' evolutions, have rather large Schwarzschild radiuses, whereas the sizes of compact objects in SMBHs centers evidently are much lesser than the $R_{gS}$.

For example, Sagittarius A* (SMBH in Milky Way) has mass, $M$=8.2×10$^{36}$ kg, and corresponding event horizon radius $R_{gS}$ =1.2×10$^{10}$ m. So average density of matter in this SMBH, $\rho = 1.1\times10^6 kg/m^3$. This density is much lesser than the density of neutron stars' matter ~10$^{17}$-10$^{18}$ kg/m$^3$, and so even if in Sagittarius A* center some big neutron star would be placed, its radius would be ~ 10$^4$ times lesser than the Sagittarius A* "event horizon".

Thus it looks as rather rational to assume, that in this case the phase of SMBH central object matter state, and any other BH's matter state, though, is the next phase after known now phases "ordinary matter", "white dwarf" and "neutron star" matters' states, and rather probably the SMBH central object is some dense composition of the corresponding the "seed's" primary particles, and of what is transformed from falling into the central object "ordinary" matter later. It looks as reasonable to suggest, that in this case some essentially uniform quark structure can be formed, which is stable because some "1-st origination quark degeneracy" – like "electron degeneracy", and "neutron degeneracy" in dwarf and neutron stars matter phases, at increasing of a BH central object's mass and pressure, next originations of quarks can appear, etc., whereas in SMBHs the space between the central object's surface and "event horizon" is filled by the accretion disk continuance, and by some other particles that have diffuse distribution; which are practically unobservable outside the horizon.

Nonetheless it looks as rather probable to propose that there don't exist some sharp border for matter in the event horizon. In the mechanics existence of "escape velocity" for some body by no means determines some limits of distances that lesser bodies in the "body's atmosphere" can move on which. If lesser body speed's value is near the escape velocity value, the body can move on practically infinite distance. It looks as rational to suggest, that that is true in the case when the escape velocity is equal to the speed of light as well. Note in this case, for example, that binding energy of an electron – and so the electron's "escape energy" – on the "event horizon surface shell" of Sagittarius A* is equal ~259 keV, what is comparable with a K-shell electron's quite non-exotic binding energy ~116 keV in Uranium atom.

If that is so, then really a "black hole", besides the accretion disk, has practically for sure, some "atmosphere" – or "hairs", which, though, are formed mostly from outer matter. Including in the observed SMBHs' jets, including the "bubbles" of Sagittarius A* SMBH it is nothing surprising – that are, rather probably, some SMBH "atmosphere's" specific details, which, as that is explained in standard cosmology, are formed mostly by magnetic fields that are formed by the accretion disk and SMBH itself.

Again - more interesting here is the problem what is the phase "black hole matter", including what fraction of the central objects' masses are constituted by the primary particles, etc.

***The "dark energy"*** is introduced in standard cosmological $\Lambda$CDM (Lambda-cold dark matter) model as the parameter that is necessary to "explain" another mysterious in mainstream physics experimentally observed effect that Matter is constantly expanding in the 3D space with different accelerations from some initial size; and the "dark energy" is some "negative energy", which (the energy), since the $\Lambda$CDM is based on the general relativity, and so the observed Matter expansion in this case happens because of/as the 3D space expansion, is some attribute of the space, and its action is evaluated by introducing of the new, specific only in cosmology, fundamental "$\Lambda$ constant" into the GR equations as some "repulsive gravity".

Really the ΛCDM is so based on fundamentally incorrect postulates – Matter's spacetime is fundamentally absolute [5]4D Euclidian spacetime, and by no means imaginary mathematically pseudo Riemannian 4D space; which fundamentally can not be transformed, including expanded, and, besides, the GR principally isn't applicable to the first "Matter's expansion" – to the exponential "expansion" in "inflation epoch", which is some – and rather probably rational – part of the ΛCDM.

Both interpretations of existent cosmological data as "space expansions", i.e. the exponential "inflation" on the first step of Creation, and more tolerant next one that rather probably really proceeds till now, really, if really happen, really are the FLE lattice expansions; and to make that it was – and is – indeed necessary to pump into the lattice essential energy. However, this energy is completely outside physics, and so attempts to incorporate this energy in existent physical theories, as that is in cosmology at the introducing of Lambda term in the GR equations, which determinates "space expansion", really are irrational.

Though, as that is in this Beginning model, some mathematical estimations of this (in the model – Creator's) energy can be used in rational purely phenomenological descriptions of what and how happened in first instants at Beginning, or, for example, when for description of the FLE lattice "2-nd" expansion after appearance of "ordinary" Matter it looks as would be possible to find some rational reasons as well. For example, this expansion would be necessary to prevent Matter's collapse because of the "gravity paradox" [Seeliger**,** 1895]; though here can be many other reasons, of course.

## 9. Discussion and conclusion

The foundational base of proposed here informational ("The Information as Absolute") concept is the rigorous proof of that everything what exists (can exist, "cannot exist") is/are nothing else than some informational patterns/systems of the patterns that are elements of absolutely infinite "Information" Set. The Set, in turn, is some unity of some set of "inert" elements and of "active" set of utmost fundamental Rules, Possibilities, Quantities, etc., "Logos" that "make information", though to separate notions "inert" and "Logos" is impossible. They are complementary, both are defined only in a unity; and the Logos set elements are informational patterns also.

More concretely the concept includes existent information and set theories (as well as all other sciences, though); in some concrete cases in similar way but not identically. Existent information theories, i.e. Hartley–Shannon's, complexity and automata theories, logics, language theories, etc., correspond only to some specific properties of the information. These properties (for example, the possibility to measure the "quantity of information" by using the values of logarithms of the probabilities of possible outcomes) rather probably correlate with some very common "rules of existence and interactions" of the elements in the Set "Information" that are hidden till now in Logos, besides these theories are rigorously formalized and developed in compliance with criteria of truth, consistency, completeness, etc. Thus the existent information and set theories, as well as the mathematics as a whole (which in reality eventually is a specific information theory), are directly involved in this concept and can be directly applied in concrete cases, an important example is investigations of the rigorously logically organized informational system "Matter", where. mathematics is the extremely effective tool.

However, "conscious" using mathematics in this case i.e. understanding – why it is the extremely effective tool, is possible only in framework of the whole concept, when it becomes to be clear, that *mathematics isn't physics*, *it is only a tool*, while Matter's construction is determined, first of all, by the set of fundamental laws/links/constants, which though can be formulated mathematically, but cannot be derived independently directly from any mathematical theory; and when in mainstream physics, in accordance with rather popular in physics the Dirac's wording "shut up and calculate", time to time corresponding physical theories appear, that are in most cases some strange constructions, which, though are mathematically perfect, are senseless in indeed physics.

.

On higher, Meta-mathematical, level of consideration the mathematics itself calls for the substantiation, though. K. Gödel defined the purview of the set theory as (quoted in [Maddy, 2005]: [if the concept of set] "…is accepted as sound, it follows that the set-theoretical concepts and theorems describe some well-determined *reality*..." The concept suggested here clarifies to certain extent what is this "well-determined *reality*", which, in fact, mathematics studies. Note, however, again, that this reality isn't the whole "Information" Set, Information is extremely bifurcated phenomenon to be completely formalized, the "absolute infinity" is equivalent of absolute uncertainty.

In contrast to mathematics, which is practically purely product of data processing in the fundamentally non-material "mind mode" of consciousness operation, where the data can be, and in mathematics are, defined arbitrarily by consciousness only, i.e. without any relations with/to what a consciousness observers in the environment, the subject domains of nature sciences, philosophical, and religious, conceptions, relate to principally outer to consciousness environment, and so in mainstream philosophy and sciences the basic postulates are fundamentally transcendent. In religions (in fact, in Idealism also) the principal impossibility of the cognition [at least by human's consciousness] of the divine design is postulated tacitly or not. Materialism, as a rule, considers this problem rather superficially, what, nonetheless, doesn't change something in the fact that the conception of the existence of some eternal Matter is absolutely equally mystical and transcendent as the conceptions of, for example, eternal God in Christianity or eternal Spirit in Hegelian philosophy.

Principal transcendence of main mainstream philosophical notions/phenomena "Matter" and "Consciousness" results till now in the fact that other sciences, which study concrete material, alive and conscious objects, really study something what is transcendent and so principally non-cognizable; and so when something is, nonetheless, cognizable adequately to objective reality in everyday practice and sciences, that happens for/by really some mysterious reasons and ways. Thus, though corresponding "cognition problems" are studied yet for a long time in framework of one of main philosophical branches "Epistemology", these problems really remain be solved only as formulation of instinctively known by humans of some really transcendent certain criteria, rules, ways at studying of principally transcendent things, which are adequate to objective reality; and these epistemological findings are practically equally instinctively known also for all/every living beings, which also instinctively and adequately apply these criteria, rules, ways in their everyday practice (more see sections 7.2., 7.3.).

Though the rather evident transcendence of the mainstream philosophical doctrines, schools, etc., isn't an obstacle for existence of the philosophy as legitimate science, where, as in "The Problems of Philosophy" [Russell, 1912: ch. 2] B. Russell [Russell, 1971] wrote: "…but whoever wishes to become a philosopher must learn not to be frightened by absurdities…"; and most of philosophers are rather bold people.

Nature scientists don't pay some attention to the epistemological problems, really instinctively studying Nature, and in many cases make that rather successfully, because of that Information, despite of is absolutely fundamental phenomenon, isn't transcendent, whereas the cognizing consciousness is made from the same stuff "Information" as everything else. *Thus there is fundamentally nothing mysterious in that some informational system, which is able to obtain and logically analyze the obtained, information in some case makes that adequately to the objective reality; and, however principally in some cases makes that illusorily.*

Though we must note, that the last statement will be completely true only after the Logos element "Energy" will be really scientifically defined, it remains till now mysterious, and its action is essentially understandable only *at application* to the Set's simple binary reversible element "Matter".

Including the fundamental for humans now Meta-notions/phenomena "Matter" and "Consciousness", including "consciousness on Earth", become be rather rationally defined, discerned and cognizable as some informational systems. As well as, though, any principally transcendent philosophical and religious postulates and "designs" turn out to be principally cognizable also; when, say, a God or Spirit now must not be omnipotent to create Universe from nothing; that are, if exist, nothing else than some informational systems in the Set. In turn, studying the Set's properties, Materialism obtains some possibility to study rationally materialistic versions of the Matter's Beginning and evolution, etc., though from this concept rather rationally follows that these materialistic versions are impossible with a well non-zero probability, however this probability isn't equal to 1 for sure. Though this problem now doesn't look as a purely philosophical problem, and really that could be a Meta-problem in physics, however that cannot be a physical problem, since in physics it hasn't some concrete solution principally – physics never will know what is in Matter below Planck scale.

.

Note, though, that the fundamentally mainstream philosophical problems of ontology "Matter" and of "consciousness on Earth", now become to be concrete fundamental problems in concrete sciences, "Epistemology" becomes to be some superfluous philosophical branch, real cognitive problems, validity criteria, etc., in every concrete case at real researches are known for real researchers much better than for philosophers,

- whereas in the real philosophy the other, and really utmost fundamental, subjects for study are posed in the concept – the absolutely fundamental phenomena/notions "Information" and "Information" Set, and, what is utmost fundamental and actual for humans – "consciousness on Earth".

In the conception till now we cannot exclude a tendency to self-organization for (at least of some) subsets that are singled out by a certain way in the Set, is inwardly inherent to the information (see also section 7.3); and, if a self–organization is an intrinsic property of information, then the Set as a whole can be, in principle, classified as some "Prime Creator", "Deo", as, e.g., G. Cantor said (quoted in Wikipedia):

> "…The actual infinite arises in three contexts: first when it is realized in the most complete form, in a fully independent otherworldly being, *in Deo*, where I call it the Absolute Infinite or simply Absolute…"

But, on another hand, here a problem appears – can we consider an Essence as some independent "Prime Creator", when this Essence *is always absolutely complete* and so cannot change anything in Himself? Insofar as even the Essence will attempt to change something, for example to begin evolving in the Set of an Universe, He must absolutely exactly follow the scenario of this change, since this scenario exists "always", including "absolutely far before" the Beginning.

This concept also allows studying on a higher level of understanding the problems in natural sciences. An epistemological example was mentioned above, i.e., the problem of remarkable adequacy of languages of scientific theories, first of all mathematics, in describing and analyzing of material objects and their interactions, however, since in it practically all fundamental phenomena/notions in concrete sciences are scientifically defined, the concept turns out to be the concrete base of concrete sciences. Now that resulted in development of *the informational physical model of Matter* [Shevchenko, Tokarevsky, 2021, 2022], where more 30 fundamental problems in physics, including in cosmology, are either solved or clarified on the level when possible rational ways of solutions have become to be rather certain.

The fundamental, and utmost actual for humans, "consciousness problem" is principally, and essentially concretely, clarified as well [Shevchenko, Tokarevsky 2018a] in the first approximation functional consciousness model, where the relations in the systems "Matter ⇔ Alive ⇔ Consciousness" also seems more understandable, including the problem "what is Life", where in the concept it is rigorously proven that any consciousness in the Set, including the "consciousness on Earth" version, is fundamentally non-material, however evidently interacts, as that evidently follows from everyday facts, when conscious actions transform into a material action, for example when a human's consciousness controls his material body by some non-material "unphysical" forces.

Thus at least primary physical-chemical processes, resulting in creation of some protein macromolecules and DNA (RNA), could be controlled by some primitive *non-material* informational structure, which rather probably developed eventually in the human's consciousness and which built at that to herself a stable house, a body. At that seems as rather probable that the consciousness on recent stage of development is organized similarly to usual computer, i.e. the consciousness consists of a "power supply" (human's body), "mother board" and "hard disk" (brain, which operates as the long term memory), "random access memory" ("short term memory" in psychology), "processor", and "BIOS". The last three functional modules are utmost probably just non-material and exist/operate outside Matter in the consciousness's spacetime, which is essentially doesn't overlap with Matter's spacetime; and both determinate basic consciousness's functions.

Note also, that despite of that "ordinary" computers aren't purely material structures since are designed by non-material consciousnesses and are governed by purely non-material products of the consciousnesses "program shells", and some computers well imitate functions and behaviour of real consciousnesses, really any computer cannot be some consciousness, "AI" fundamentally cannot be "I",

- nonetheless we cannot exclude, that in computer evolution some sensors, sensible enough to be controlled by consciousness surely including directly by thoughts, will be developed. Then a variant is possible when human consciousness someday will move to a new residence, obtaining a stable and well reparable iron organism which does not require for its existence biological food, etc. Though such "humans" will, possibly, with greater pleasure drink benzine that was seasoned, say, 40 years in an oak tun.

Above we considered mainly ontological and epistemological aspects that relate, first of all, to Nature sciences, but the concept can be applied in humanitarian domains of philosophy also. Here seems worthwhile to make some remarks relating to main existent idealistic and materialistic doctrines that consider ways of future evolution of the humanity. The idealistic (first of all, religious) ones usually consider this evolution as fatally controlled by some mighty transcendent external forces. Materialistic, first of all Adam Smith concept, and Marxism [Marx, Engels, 1848], which in this part is based on this concept, doctrines contend that the evolution is controlled practically completely by some "materialistic interest" ("historical materialism") and corresponding economical laws; in every society "the *material Being controls the [individual and social] Consciousness*".

Really the indeed well observed the "materialistic interest" action in individual and social behavior of humans is the actualization of the human's consciousness program shell set of resident utilities that govern the consciousnesses – and so humans – aimed at providing stability of the consciousness's practically material residence - body, i.e. satisfactory of the body's needs. The needs are rather far from the inward consciousness's non-material needs, nonetheless these utilities process rather well determined and concrete information, which rather stubbornly and convincingly enters from the practically material sensors – in full contrast to the practically completely unclear and uncertain for most of humans now non-material needs, and so yeah, material interest governed humans and societies in whole history of humanity, and till now; bizarrely shuffling fates of humans and countries.

Correspondingly Marxism (see also [Shevchenko, Tokarevsky, 2018b]), as that rather convincingly looks, turned out to be a really essentially scientific doctrine that till now adequately explains the social evolution on passed historical period ("primitive communism – capitalism" succession of social systems), however really this philosophical doctrine, in contrast to Adam Smith concept, contains also first of all some formulation of some really non-material consciousness needs, which were formulated earlier in some, first of all Utopian, "socialistic" social models.

The models were some really natural continuation of the process, when because of technological development in last few thousands of years production of material wealth and corresponding social relations developed so, that the constant priority of material utilities was weakened, and so the non-material needs become rather actual in humans. That resulted in essential "humanization" of some, so advanced and widely expanded, religions, and developing of models of societies, where the "spiritual" needs and values would be more significant – mostly of some "socialisms" above, which, though, in this case mostly contained rational versions of civil implementation of "humanitarian" religious dogmas in recent (in those times) societies; a number of attempts of practical realizations of the models failed.

So Marxism really isn't completely materialistic theory, and included most of socialistic models principles, which really limit the material interest utilities action, however, since for authors of Marxism (and for authors of the socialistic models also, though) the phenomenon "consciousness" had sense only as some principally transcendent ephemeral, and in Marxism fundamentally non-existent, religious "soul", explains the socialistic models' fails, and formulates – as that states – the really rational ways of the implementation, makes that specifically materialistically

– it states that the "asocial" actions of material interest, i.e., first of all unfair unmerited distribution of produced material wealth, and other evidently wrong traits, in capitalism and earlier social formations, are caused by elemental mode of production and distribution of deficit material wealth, which are deficit because of that the modes of production are elemental, and if the production of material wealth will be planned, then "the wealth will flow whole flood", and in such society the "Great Principle" of the corresponding social formation "communism" [Marx, "Critique of the Gotha Program"] *"From each according to his ability, to each according to his needs*" will be realized, and so all non-material problems in societies thus will be solved.

.

In XX century a number countries attempted to implement the main postulates of Marxism, however this communistic experiment has ended in full contradiction with historical materialism, and really the first socialistic country, USSR, in fact has evolved after socialistic 1917 year revolution only through all historical materialism social formations – from primitive "military communism" through really some "socialistic" slavery and feudalism, and inevitably logically eventually into firstly wild and recent capitalism. When material interest really governs humans and social relations, principally nothing besides capitalism can exist in technologically developed enough societies.

From that, however, it doesn't follow that now humanity is on some final stage of its social development, and, at that it looks as rather probable that the real development will proceed in accordance with the principle "*In all/every humans' societies individual and social Consciousness determinates the material individual and social Being*".

This concept now doesn't consider some concrete traits, social structures, etc., of corresponding so post-capitalistic societies, and only note here some points, which look as rather essential at elaboration of this problem. First of all, that is the rather evidently and convincingly observed *trend in living beings development in* *past couple billions of years* "*more and more outside Matter into other Set's regions*", from the first biostructures up to the "homo sapiens sapiens" version, pointed above. It looks as rather rational to conjecture that this trend, which evidently is as some realization of non-material consciousness's needs and aims, will continue even in the background "subconscious" mode, as that was all existence of Life on Earth, when this trend realization looks as some way of "trials and errors";

- but now it looks as rather rational to conjecture that this trend can be made as more rational, for example yet now science, first of all biosciences, are developed up to the level, when some changes in human's material structures aimed at developing some functional consciousness abilities at obtaining and processing in mind mode of operation could be enhanced, etc.

Though we cannot exclude case that the consciousness's development scenario with rather non-zero probability has not finished on "homo sapiens sapiens" version, and it contains sequels, where the greater complexity of the consciousness and her capabilities to apprehend new information and to control more and more regions/ structures in the Set will be realized, as the observable now development "First biostructures Consciousness" → "Alive Consciousnesses → "human Consciousness" will continue as "…human Consciousness → "[human?] Consciousness-1" → "Consciousness-2"…; where "Consciousness-N" mean next versions of now "human Consciousness" in the Set, which will be based on other, and probably arranged by qualitatively another way, corresponding Consciousnesses' basic elementary fundamental logical elements.

Though even in existing situation current human's material needs, though important for the consciousness, as is important for her existence, objectively can have, first of all as a result of parallel technological development, lesser and lesser role in human's life, and the technological development yet now essentially decreases humans' participation in wealth production processes, providing for humans' more and more free time, which can be spend so by individuals aimed at their spiritual development. Though that is possible really only if a human consciously enough understands that the spiritual development is just utmost natural development, and it fundamentally must be above the material interest. In other cases, the material interest will control the free time humans' actions as well, as that was observed till now, when the material interest needs are unlimited.

However, what more important, the real spiritual development is possible only if the corresponding aims, ways, and techniques of the development are really scientifically rationally defined, and just corresponding problems become to be the main subjects of just real philosophy – if it will be real science and really will seek for the scientific answers to *the ultimately important really questions – what is the place and role of "consciousness on Earth" and human's consciousness in the "Information" Set? And what are the optimal ways of the consciousness development to occur in the optimal the place and role*?

It is evident, that these problems can be solved principally essentially basing also on corresponding researches in other sciences, which in some times singled out philosophy as concretizations of problems and methods of "primary philosophy" branches "Physics" and "Logics"; and, after in this concept the remaining in this branches "too fundamental" in the sciences problems ontology of the Matter, Space, Time, etc., are solved, and so these phenomena have became subjects for rational study in the sciences, in really philosophy just philosophical branch "Ethics", as "Ethics in the "Information" Set" remained, which never be rationally quantitatively formalized, as that other sciences require, however just *this branch* *becomes to be utmost important at really scientific designation of humans' individual and social development.*

**Acknowledgments**

Authors are very grateful to Professor Brodin M.S., Institute of Physics of NAS of Ukraine, for useful discussions and support, and to Dr. Shevchenko A.S., I.I. Schmalhausen Institute of Zoology of NAS of Ukraine for the initiation of this concept's development.